\pdfoutput=1
\documentclass[useAMS,usenatbib]{mn2e}
\usepackage{natbib}
\usepackage{color}
\usepackage{amsmath,amssymb}
\usepackage{graphicx}
\usepackage{bm}

\newenvironment{mytabular}[1][1]{
  \renewcommand*{\arraystretch}{#1}
  \tabular
}{
  \endtabular
}

\usepackage{color}

\newcommand{\de}{\mathrm d}

\title[Estimating rotation in radio cosmic shear surveys]{Estimating the weak-lensing rotation signal in radio cosmic shear surveys}
\author[D. B. Thomas et al. 2016] {Daniel B.~Thomas$^{1,2}$\thanks{E-mail: thomas.daniel@ucy.ac.cy}, Lee Whittaker$^2$, Stefano Camera$^2$, Michael L.~ Brown$^2$  \\ \\ 
  1 - Department of Physics, University of Cyprus, Aglantzia, Nicosia, 2109\\
  2 - Jodrell Bank Centre for Astrophysics, School of Physics \& Astronomy, The University of Manchester, Manchester M13 9PL, UK}

\date{Accepted ;
  Received ; in original form }
\pagerange{\pageref{firstpage}--\pageref{lastpage}} \pubyear{2016}

\begin{document}

\maketitle

\label{firstpage}

\begin{abstract}
Weak lensing has become an increasingly important tool in cosmology and the use of galaxy shapes to measure cosmic shear has become routine. The weak-lensing distortion tensor contains two other effects in addition to the two components of shear: the convergence and rotation. The rotation mode is not measurable using the standard cosmic shear estimators based on galaxy shapes, as there is no information on the original shapes of the images before they were lensed. Due to this, no estimator has been proposed for the rotation mode in cosmological weak-lensing surveys, and the rotation mode has never been constrained. Here, we derive an estimator for this quantity, which is based on the use of radio polarisation measurements of the intrinsic position angles of galaxies. The rotation mode can be sourced by physics beyond $\Lambda$CDM, and also offers the chance to perform consistency checks of $\Lambda$CDM and of weak-lensing surveys themselves. We present simulations of this estimator and show that, for the pedagogical example of cosmic string spectra, this estimator could detect a signal that is consistent with the constraints from \emph{Planck}. We examine the connection between the rotation mode and the shear $B$-modes and thus how this estimator could help control systematics in future radio weak-lensing surveys.
\end{abstract}


\section{Introduction}
The gravitational deflection of light is one of the key results from Einstein's General Relativity (GR), and has several important consequences for cosmology. In particular, whenever we observe the sky, the objects we observe will have been distorted slightly, be they galaxies or even the Cosmic Microwave Background. This low level distortion is known as weak lensing by large scale structure, or cosmic shear.

This distortion is now used by cosmologists to provide information on the gravitational potentials and thus the matter content in the Universe. Since the first observations \citep{wittman2000,bacon2000,kaiser2000,vanwaerbeke2000}, multiple weak-lensing surveys have been carried out, leading up to the recent CFHTLens  \citep{cfhtlens} and KiDS \citep{kids450} results and the ongoing DES survey \citep{des}. The importance of weak lensing will increase further with the significantly larger upcoming surveys, including LSST \citep{lsst} and the ESA \emph{Euclid} satellite mission\footnote{http://euclid-ec.org} \citep{euclid,euclid2,euclid3}.

In the standard presentation of weak lensing, the possible distortions that are considered are the convergence ($\kappa$), a change in size of the image, and the shear ($\gamma_1$ and $\gamma_2$), which represent a change to the ellipticity of the image. For a scalar gravitational potential, this is a complete description at first order. However, if tensor or vector gravitational potentials are present, or the calculation is carried to higher order, then a rotation mode $\omega$ is also present. As the name suggests, this causes a rotation of the image.

This rotation mode is rarely studied in the literature as it is a higher order effect in the standard $\Lambda$CDM cosmology, although it has been looked at in other contexts, see e.g. \cite{pen2006}.  Furthermore, there is no estimator for the the rotation signal in cosmological weak-lensing surveys. Indeed, in standard cosmic shear surveys that measure gravitational lensing by observing galaxy shapes, measuring the rotation signal is impossible since there is no information available on the original orientation of the galaxy image (before it was lensed).

The size of the second order effect in $\Lambda$CDM was calculated in \cite{krause2010} and was found to be small. However, it is possible to generate the rotation mode to a greater magnitude in scenarios beyond $\Lambda$CDM. This requires sourcing vector and/or tensor modes in the metric at first order. One example of this in the literature is \cite{thomas2009}, where the rotation signal is calculated for a network of cosmic strings. However, due to the lack of an estimator for the rotation modes, it was considered that the signal would be found through the $B$-modes of cosmic shear. We will elaborate on the connection between rotation and $B$-modes later in this paper. Further work was carried out in \cite{yamauchi2012} and \cite{yamauchi2013} to investigate how the $B$-modes could be used to constrain cosmic string networks.

In this paper, we propose an estimator for the rotation signal. It based on extending the work of \cite{brown2011}, where the polarisation of radio galaxies is used as additional information in weak lensing surveys \citep[see also][]{whittaker2015}. The radio polarisation is used to obtain an estimate of the position angle of the unlensed galaxy image. By comparing this to the final lensed image of the galaxy, it is possible to determine whether the image has been affected by rotation in addition to shear and convergence.

We will present the estimator and then present the simulations that were used to evaluate the effectiveness of the estimator. Using the example of cosmic string power spectra, we show that this estimator could detect a signal that is consistent with the constraints from \emph{Planck} \citep{planckstrings}. We also discuss how this estimator can be used to help control systematics in weak-lensing surveys.

The notation we will use is as follows: upper case roman letters denote indices that run over 1-2, i.e. the directions transverse to the line of sight. Lower case roman letters denote indices that run over the three spatial dimensions, and Greek letters denote indices that run over all four spacetime dimensions. The comoving distance will be denoted $\chi$ and our transverse coordinates are expressed as a (small) angle multiplying the comoving distance $\chi$, such that $x^{i}=\chi(\theta^1,\theta^2,1)$.

In section \ref{sec_wlrot} we recap the rotation signal in weak lensing and then review radio weak lensing in section \ref{sec_radiowl}. In section \ref{sec_estimator} we present our rotation estimator as well as some properties of the estimator. In sections \ref{sec_cosmosims} and \ref{sec_results} we present the details of the simulations that we performed to test the estimator and the results of those simulations. We conclude in section \ref{sec_conc}.

\section{Rotation in weak lensing}
\label{sec_wlrot}
In the standard picture of weak lensing, the $2\times2$ distortion tensor $\psi_{IJ}$ (consisting of the magnification matrix $A_{IJ}$ with the identity subtracted) enumerates the possible effects that lensing can have on an image
\begin{equation}
A_{IJ}-I_{IJ}=\psi_{IJ}=\left(
\begin{array}{cc}
-\kappa-\gamma_1  & -\gamma_2+\omega \\
-\gamma_2-\omega  &-\kappa+\gamma_1 \\
\end{array}
\right)\rm{.}
\end{equation}
The four possible effects on the image are the convergence ($\kappa$), the two components of shear ($\gamma_1$ and $\gamma_2$) and the rotation ($\omega$). Estimators exist for the convergence and shear, but thus far no estimator has been proposed for the rotation mode. It is simple to see why this is the case. Shear estimators are statistical, in the sense that they detect correlations between galaxy orientations, and thus detect a coherent distortion. Convergence estimators are also statistical, and detect a coherent deviation of the sizes from the average. For the rotation, there is no ``average'' position angle to detect a coherent deviation from. Furthermore, given a set of galaxy position angles, applying a rotation to all of the galaxies creates an entirely equivalent set, since there is no preferred direction on the sky. The information that would be needed in order to detect rotation is the intrinsic position angles of the galaxies, i.e. their orientation before their images are lensed. The estimator proposed in this paper uses polarisation of radio galaxies in order to estimate the intrinsic position angles. The estimator will be presented in detail in section \ref{sec_estimator}.\\

To first order in image distortions, $\omega$ as defined above corresponds to a clockwise rotation of the image by an angle $\phi\approx \omega$, measured in radians. To see this, consider the action of a rotation matrix $\mathbf{R}$ for a small angle $\phi$, on a magnification matrix with no rotation
\begin{eqnarray}
\mathbf{R}&=&\left(
\begin{array}{cc}
\cos \phi  & \sin \phi \\
-\sin \phi  & \cos \phi \\
\end{array}
\right)\approx\left(
\begin{array}{cc}
1  & \phi \\
-\phi  & 1 \\
\end{array}
\right)\\
\mathbf{A'}
&=&\left(
\begin{array}{cc}
-\kappa'-\gamma_1'  & -\gamma_2'+\omega' \\
-\gamma_2'-\omega'  &-\kappa'+\gamma_1' \\
\end{array}
\right)\nonumber\\
\mathbf{A}
&=&\left(
\begin{array}{cc}
-\kappa-\gamma_1  & -\gamma_2\\
-\gamma_2 &-\kappa+\gamma_1 \\
\end{array}
\right)\nonumber\\
\mathbf{A'}\hspace{-0.2cm}&=&\hspace{-0.2cm}\mathbf{R}\mathbf{A}=\left(\hspace{-0.1cm}
\begin{array}{cc}
-\kappa-\gamma_1-\phi\gamma_2  & -\gamma_2+\phi(1-\kappa+\gamma_1) \\
-\gamma_2-\phi(1-\kappa-\gamma_1)  & -\kappa+\gamma_1+\phi\gamma_2 \\
\end{array} \hspace{-0.1cm}\right)\nonumber
\text{.}
\end{eqnarray}
Thus, an anti-symmetric component $\omega'=\phi(1-\kappa)\approx\phi$ is generated. The shear components transform as $\gamma_1'=\gamma_1+\phi\gamma_2\approx\gamma_1$ and  $\gamma_2'=\gamma_2-\phi\gamma_1\approx\gamma_2$, so to first order in image distortions, the rotation $\phi$ corresponds to the anti-symmetric component $\omega$ and leaves the shear and convergence unaffected. The image amplification is given by det($\mathbf{A}$), to which the rotation contributes $\omega^2$. Again, since we are dealing with small distortions, this factor is ignored in the same way that the contribution of the shear to det($\mathbf{A}$) is typically ignored. \\

More generally,  $\phi=\arctan \omega$. To see this, we can split the rotation matrix into a change to the size of the image ($S$) plus an anti-symmetric matrix
\begin{equation}
\left(
\begin{array}{cc}
\cos \phi  & \sin \phi \\
-\sin \phi  & \cos \phi \\
\end{array}
\right)=S\left(
\begin{array}{cc}
1  & \omega \\
-\omega  & 1 \\
\end{array}
\right)\text{.}
\end{equation}
Equating the two sides, we get $S=\cos\phi$ and $S\omega=\sin\phi$, thus $\arctan\omega=\phi$ as expected. 

\subsection{Rotation from cosmological perturbations}
Here we will show a simple example of how the rotation mode can be generated by cosmological perturbations. The perturbed FLRW metric in Poisson gauge is given by
\begin{align}
&g_{00}=-1-2\Psi\\
&g_{0i}=-aV_i\\
&g_{ij}=a^2\delta_{ij}(1+2\Phi)+h_{ij}\text{,}
\end{align}
where $\Psi$ and $\Phi$ are scalar perturbations, $V_i$ is the divergenceless vector perturbation and $h_{ij}$ is the transverse traceless tensor perturbation. Scalar perturbations do not generate rotation at first order due to the symmetry between $\psi_{12}$ and $\psi_{21}$ when generated only by scalar perturbations. In a $\Lambda$CDM cosmology, on linear scales, the vectors and tensors are not generated at first order. Furthermore, even on non-linear scales in $\Lambda$CDM, the vector and tensor perturbations are negligible \citep{bruni2014, adamek2014}. At second order however, the scalar perturbations can generate rotation, through the term in the deflection angle that corresponds to the Born correction and lens-lens coupling \citep[see][]{dodelson2005,krause2010,thomas2015}. The size of this effect (and equivalently the size of the $B$-mode, see later) at second order in $\Lambda$CDM cosmologies has been examined in \cite{krause2010} and \cite{jain2000} and found to be orders of magnitude smaller than the leading order signal. This is one of the reasons that rotation has received little attention in the literature. However, vector and tensor perturbations can both source the rotation component at first order. For example, for the vector perturbation in the above, the power spectrum of the rotation can be expressed explicitly as \citep[see e.g.][]{thomas2015,thomas2009}
\begin{eqnarray}
\frac{1}{4\pi^2}C^{\omega}_{\ell}=P_{\omega}(l)=\frac{1}{4}\int^{\chi_{\infty}}_0 d\chi g^2(\chi)\frac{a^2l^2}{\chi^2}P_{\dot{V}}(l/\chi)\text{,}\nonumber\\
\text{with} \left<\dot{V}_i \dot{V}_j\right>=\left(2\pi\right)^3\delta(\vec{k}-\vec{k}')(\delta_{ij}-\hat{k}_i\hat{k}_j)P_{\dot{V}}(l/\chi)\text{.}
\end{eqnarray}
Here, $\chi$ is the comoving distance, $\chi_{\infty}$ is the maximum comoving distance to a source in the survey, $\dot{V}$ is the time derivative of the vector potential and we have used the Limber approximation and expressed the galaxy distribution and standard weight function as $g(\chi)$ as in \cite{dodelsonbook}. Note that these calculations have been performed using the displacement approximation, we comment further on this below. A similar equation holds for the tensor perturbation.

Thus, any extensions of $\Lambda$CDM that generate vector and/or tensor perturbations at first order may generate a non-negligible rotation signal. In this sense, this signal acts as a null-test of the $\Lambda$CDM cosmology.  A detection of the rotation mode would either mean that an extension to $\Lambda$CDM is required or that the weak-lensing survey systematics were not understood. To examine the sensitivity of our proposed estimator (see section \ref{sec_estimator}), we will need to use a specific model. The example that we will use in sections \ref{sec_cosmosims}-\ref{sec_results} is a network of cosmic strings.

The rotation mode offers a further consistency check beyond just whether it is detected. On all but the largest angular scales it should be equivalent to the $B$-mode of cosmic shear (see section \ref{sec_bmoderot}). Thus, if the two are detected and not equal, then there is either a systematic error with the weak-lensing survey or one or more of the assumptions or approximations underlying the weak-lensing calculations is incorrect.\footnote{There is a third possibility, which is that the difference is being generated by tensor perturbations. For these, due to the metric shear term \citep{dodelson2003,schmidt2012}, the displacement approximation and thus the rotation to $B$-mode correspondence does not hold.} Thus we can see that attempting to detect the rotation signal is a fruitful endeavour, particularly as it will essentially come for free in a radio survey. 

\subsection{Rotation to $B$-mode correspondence}\label{sec_bmoderot}
We will now recap why the rotation and $B$-mode signals are equal. As mentioned earlier, the calculations referenced earlier use the displacement approximation. This is the statement that the weak-lensing effects can be calculated as a deflection from a ``straight path'' in an underlying flat geometry \citep{stebbins1996}. In terms of the calculation, this approximation means that the distortion tensor describing the effects on images at a point in space can be expressed as the Jacobian of the weak lensing deflection angle $\alpha$. Typically, the deflection angle is then calculated using the geodesic equation.

Assuming the displacement approximation holds, it is possible to write down simple relations between the shear, convergence and rotation. This is because the displacement field only has two degrees of freedom, the scalar and pseudo-scalar (or curl) potentials \citep[e.g.][]{hirata2003,stebbins1996}. Here, we will use the complex notation for the shear \citep[see e.g.][]{bacon2006}
\begin{align}
&\alpha=\alpha_1+i\alpha_2\nonumber\\
&\gamma=\gamma_1+i\gamma_2\nonumber\\
&\partial=\partial_1+i\partial_2\nonumber\\
&Z=\kappa+i\omega\nonumber\\
&L=l_1+il_2\nonumber\\
&C=E+iB\nonumber\\
&A^*=A_1-iA_2\hspace{0.5cm}\rm{for}\hspace{0.5cm} A\in\{\alpha, \gamma, \partial, Z, L, C\}\text{.}
\end{align}
Here, $\alpha$ is the deflection angle, $\partial_1$ and $\partial_2$ are the derivatives with respect to $x_1$ and $x_2$, the two components of the 2D coordinate system on the sky, and $\alpha_1$ and $\alpha_2$ are the two components of the displacement. The $E$- and $B$-modes will be defined shortly and $l_1$ and $l_2$ are the Fourier conjugates to $x_1$ and $x_2$. Under the displacement approximation, the distortion tensor given earlier is given by the Jacobian of the deflection angle, $\psi_{IJ}=\alpha_{I,J}$, where ``$_{,J}$'' denotes the derivative with respect to $x^J$. Expressing the convergence, rotation and two components of shear in terms of the components of the distortion tensor, and therefore $\alpha$,
\begin{align}
 Z=\frac{1}{2}\partial^*\alpha\nonumber\\
\gamma=\frac{1}{2}\partial\alpha\nonumber\\
\partial Z=\partial^*\gamma \rm{,}
\end{align}
so now the relationship between the convergence, shear and rotation is manifest. We wish to convert the shear into the $E$- and $B$-modes, so we move to Fourier space, $\partial_i=-j l_i$\footnote{We use $j$ to avoid the confusion between the imaginary number $i$ in the definitions of the weak lensing quantities and the imaginary number $i$ (now $j$) in the Fourier transform.}. Notice from the definitions above that  that $-jL=\partial$ and $LL^*=l^2_1+l^2_2=l^2$, so $\partial Z=\partial^*\gamma$ becomes $L Z=L^* \gamma$. The definitions of the $E$-and $B$-modes are (see e.g. \cite{dodelsonbook})
\begin{align}
E=\frac{l^2_1-l^2_2}{l^2}\gamma_1+\frac{2l_1l_2}{l^2}\gamma_2\nonumber\\
B=-\frac{2l_1l_2}{l^2}\gamma_1+\frac{l^2_1-l^2_2}{l^2}\gamma_2 \rm{.}
\end{align}
It then follows that $C=\frac{(L^*)^2}{l^2}\gamma$. We can substitute the definition for $Z$ into this equation and use the above relationship between $Z$ and the shear to find
\begin{equation}
 C=\frac{(L^*)^2}{l^2}\gamma=\frac{LL^*}{l^2}Z=Z\rm{.}
\end{equation}
Thus, simply working from the definition of the distortion tensor and the displacement approximation, we can understand why the rotation field maps to the $B$-mode of shear and the convergence field maps to the $E$-mode of shear.

We note that these relationships break down if the displacement approximation breaks down. Although there are good reasons to believe that the displacement approximation holds, see \cite{stebbins1996} for a nice discussion of these, there are circumstances in which it is known to break down. Firstly, it is known that it does not hold for tensor perturbations, this is the origin of the ``metric shear'' term \citep[see e.g.][]{dodelson2003,schmidt2012}. This is not an issue in this work as we do not deal with tensor perturbations. However, it does mean that the rotation $B$-mode correspondence will not hold for tensor perturbations and hence that detecting non-equal rotation and $B$-modes on small angular scales could imply the existence of non-negligible tensor perturbations. Secondly, the displacement approximation is known to formally break down on large angular scales: \cite{bernardeau2010} have shown that the expression for the convergence cannot be written as the Jacobian of the deflection angle due to the presence of additional terms. However, these terms have not been calculated numerically and may yet be sufficiently small that the displacement approximation can still be used on large angular scales, allowing the convergence $E$-mode correspondence to be recovered. It is expected that a full sky treatment as in \cite{bernardeau2010} where the vector perturbations are treated at first order would lead to terms that could contribute to the rotation mode at large angular scales, thus potentially breaking the rotation-$B$-mode correspondence.

Note that in \cite{yamauchi2012}, it is stated that there is no rotation at first order despite the metric having arbitrary metric perturbations. This statement is obviously in conflict with the explicit calculation above. However, we note that in the geodesic deviation approach \citep[as used in][]{yamauchi2012}, the basis is parallel transported along the light ray. Thus, the basis will rotate along with the image, so no rotation can be detected. This was first pointed out in \cite{stebbins1996}.

\section{Review of radio weak lensing}
\label{sec_radiowl}
Thus far, weak lensing has been predominantly studied using optical surveys. In particular, studies in the radio band have been limited due to the low number density of background galaxies. However, see \cite{chang2004} for a statistical detection of cosmic shear in the FIRST survey \citep{becker1995}. See \cite{patel2010} for an investigation of measuring the lensing signal in data from the VLA and MERLIN, where the benefits of cross-correlating radio and optical shear estimates from the same patch of the sky is also discussed. In addition, see \cite{demetroullas2015} for a demonstration of cross-correlation techniques on real data.

Increasing attention is being given to the possibility of performing weak lensing surveys in the radio band in the build up to the Square Kilometre Array (SKA)\footnote{https://www.skatelescope.org/} \citep{brown2015}. Near and mid-term facilities such as e-MERLIN\footnote{http://www.e-merlin.ac.uk/}, LOFAR \citep{morganti2010}, MeerKAT \citep{booth2009} and ASKAP \citep{johnston2008} will start to achieve a similar background galaxy number density to optical surveys.

There are several advantages to considering radio weak lensing. One is that different redshift ranges can be probed, which is particularly pronounced for the SKA. Another is that there is some additional information available in radio surveys, which can be used to help with weak-lensing systematics such as intrinsic alignments, see \cite{kirk2015} for a review of intrinsic alignments. This includes the use of rotational velocity information from HI observations \citep{morales2006} and polarisation information \citep{brown2011, whittaker2015}, more on the latter below. We will not consider instrumental and atmospherical systematics in this paper, however we note that in radio surveys some systematics are better and some are worse than in optical surveys. In addition, there is the possibility in the future to cross correlate radio and optical weak-lensing surveys \citep{camera2016,harrison2016}, which should further reduce the impact of systematics from the telescope and survey design.

\subsection{Use of polarisation of radio galaxies}
The use of radio polarisation information to improve weak-lensing surveys was first proposed in \cite{brown2011}, where it was proposed as a technique to reduce the impact of intrinsic alignments. The idea was further refined in \cite{whittaker2015}, and we will summarise this here.

Star forming galaxies are expected to dominate the observed sources in future deep radio surveys. The dominant source of radio emission from such galaxies is expected to be synchrotron radiation emitted as electrons are accelerated by large scale magnetic fields within the galaxy. This gives rise to a net polarisation position angle, which is on average anti-aligned with the plane of the galaxy \citep{stil2009}, therefore providing information about the galaxy's intrinsic orientation. Thus, we can use the polarisation information contained in radio emission of a galaxy as a tracer of the intrinsic position angle of that galaxy. Note that this polarisation information comes for free in a radio survey. One of the promising aspects of using the polarisation as a tracer of the intrinsic position angle is that this estimator is unlikely to be biased; it is implausible that any of the possible sources of scatter would prefer a particular direction. However, we note that using the polarisation information will require understanding any effects on the polarisation from intervening cosmological magnetic fields and the effects of Faraday rotation in the sources themselves as well as in our galaxy. Information from ongoing surveys should help with understanding and removing this. In particular it is expected that we can correct for this effect by extracting the rotation measures of source galaxies using information from multiple frequencies, since Faraday rotation is frequency dependent.

For a discussion of the fraction of galaxies for which we can expect to have reliable polarisation information, see \cite{brown2011}. Here we just note that present radio surveys, such as SuperCLASS, using the JVLA and e-MERLIN arrays, should provide information about the fraction of galaxies with reliable polarisation information and the error on the measurement of the intrinsic position angles from the polarisation data, $\alpha_\text{rms}$. This error can be sourced by several effects: it could come from an error in the measurement of the polarisation angle or, as already mentioned, from not correcting properly for Faraday rotation. In addition, there might be some intrinsic dispersion in the polarisation to intrinsic position-angle relationship, which would also manifest as part of $\alpha_\text{rms}$. For the purpose of this paper, the different sources of error do not matter, and will be treated together as $\alpha_\text{rms}$.

Until now, the rotation we have been discussing in this paper is the rotation of an image, i.e. an effective rotation of an ensemble of photons caused by the different deflection of neighbouring photon trajectories. One important assumption is that the polarisation is not affected by the lensing, i.e. that the polarisation angle of a single photon is not rotated. It is well established that scalar-type perturbations to the metric do not affect the polarisation angle of photons \citep[e.g.][]{dyer1992}. The case of gravitational waves is discussed in \cite{surpi1999}, \cite{prasanna2002}, \cite{kronberg1991} and \cite{faraoni2008}, with the conclusion that  their only effect is as a boundary term. In appendix \ref{app_faraoni} we comment on how the calculation from \cite{faraoni2008} can be extended to include vector perturbations, with the result that vector perturbations also only affect the polarisation angle through a boundary term. This will be far smaller than any integrated effect causing image rotation, so will be neglected for the rest of this work.

To clarify this picture further, consider the ensemble of photons making up the image of a distant galaxy. The photons trajectories will be deflected by the scalar, tensor and vector perturbations. This will cause each photon to arrive from a slightly different direction compared to an unperturbed universe, thus the image will be distorted. However, the polarisation angle associated with each photon will not change, despite the photon arriving from a different direction. Thus, the total integrated polarisation over the galaxy image will be the same as in an unperturbed universe, and therefore the polarisation can be used as an estimate of the intrinsic position angle of the galaxy.

We note that the discussion above considers Faraday rotation and lensing effects. There are other possible effects that could disturb the relationship between the polarisation angle of the galaxy and its intrinsic position angle. One effect that has received attention in the past is bi-refringence \citep[see e.g.][]{1011.4865,1411.6287}. Bi-refringence causes a rotation of the polarisation angle of the photons, which would bias our estimator in the presence of a lensing rotation signal. However, in a $\Lambda$CDM universe where the lensing rotation signal is small, our estimator would thus act as a null test for the presence of bi-refringence.

\subsection{Shear only estimator}
We will now present the use of the polarisation information when estimating the shear more quantitatively. The observed components of the ellipticity for  single galaxy can be expressed as
\begin{align}\label{eq:obs_ellip}
\epsilon_1^{\mathrm{obs}}=&\left|\bm{\epsilon}^{\mathrm{int}}\right|\cos\left(2\alpha^{\mathrm{int}}\right)+\gamma_1\nonumber\\
\epsilon_2^{\mathrm{obs}}=&\left|\bm{\epsilon}^{\mathrm{int}}\right|\sin\left(2\alpha^{\mathrm{int}}\right)+\gamma_2\text{,}
\end{align}
where $\gamma_{1,2}$ are the components of the shear induced by lensing and we have ignored any measurement error on the ellipticity. The polarisation information from the radio survey gives an estimate of the intrinsic (i.e. pre-lensing) position angle $\alpha^{\mathrm{int}}$, denoted $\hat{\alpha}^\text{int}$.

The $\chi^2$ defined in \cite{brown2011} is given by
\begin{align}\label{eq:chi2}
\chi^2=&\sum_{i=1}^Nw_i\biggl[\left(\epsilon_1^{\mathrm{obs},(i)}-\gamma_1\right)\sin\left(2\hat{\alpha}_i^{\mathrm{int}}\right)\nonumber\\
&-\left(\epsilon_2^{\mathrm{obs},(i)}-\gamma_2\right)\cos\left(2\hat{\alpha}_i^{\mathrm{int}}\right)\biggr]^2\text{,}
\end{align}
where the sum is over the $N$ galaxies in a cell on the sky (which we will refer to as sky ``pixels'') and the $w_i$ are arbitrary weights assigned to each galaxy in the pixel; these will all be set equal for the remainder of this paper. The original BB estimator \citep{brown2011} was derived by minimising this $\chi^2$, where it was shown that this estimator reduces the bias caused by intrinsic alignments. This estimator was further improved on in \cite{whittaker2015}, where the following estimator was proposed
\begin{equation}\label{eq:cBB}
\hat{\bm{\gamma}}=\mathbf{D}^{-1}\bm{h},
\end{equation}
where $\mathbf{D}$ is a $2\times2$ matrix
\begin{equation}\label{eq:D_matrix}
\mathbf{D}=\sum_{i=1}^N\mathbf{M}_i,
\end{equation}
and $\boldsymbol{h}$ is a 2-component vector
\begin{equation}\label{eq:h_matrix}
\bm{h}=\sum_{i=1}^N\mathbf{M}_i\bm{\epsilon}_i^{\mathrm{obs}}.
\end{equation}
The matrix $\mathbf{M}_i$ is given by
\begin{equation}\label{eq:matrix_M_corr}
\mathbf{M}_i=\left(
\begin{array}{cc}
\beta_4^{\mathrm{int}} - \cos\left(4\hat{\alpha}_{i}^{\mathrm{int}}\right) & -\sin\left(4\hat{\alpha}_{i}^{\mathrm{int}}\right) \\
-\sin\left(4\hat{\alpha}_{i}^{\mathrm{int}}\right) & \beta_4^{\mathrm{int}} + \cos\left(4\hat{\alpha}_{i}^{\mathrm{int}}\right)
\end{array} \right),
\end{equation}
where $\beta_4=\left<\cos\left(4\delta\alpha^{\mathrm{int}}\right)\right>$ is an average over the measurement error $\delta\alpha^{\mathrm{int}}$ on the position angle. For Gaussian errors, this simplifies to $\beta_4=\exp\left( -8\alpha^2_\text{rms}\right)$, where $\alpha_\text{rms}$ is the standard deviation of the errors. The $\beta_4$ terms correct for the noise bias coming from the angle averages. This estimator was shown in \cite{whittaker2015} to reduce the bias from intrinsic alignments to negligible levels, even in the presence of errors on the intrinsic position angle estimate. This estimator will henceforth be referred to as the corrected BB (cBB) estimator.

\section{Estimating the shear and rotation}
\label{sec_estimator}
We now present our estimator for the rotation, extending the shear only estimator from the previous section. Allowing for a rotation angle, $\alpha^{\mathrm{rot}}$, the observed ellipticity (\ref{eq:obs_ellip}) should now be written as
\begin{align}\label{eq:obs_ellip2}
\epsilon_1^{\mathrm{obs}}=&\left|\bm{\epsilon}^{\mathrm{int}}\right|\cos\left(2\alpha^{\mathrm{int}}+2\alpha^{\mathrm{rot}}\right)+\gamma_1+\delta\epsilon_1,\nonumber\\
\epsilon_2^{\mathrm{obs}}=&\left|\bm{\epsilon}^{\mathrm{int}}\right|\sin\left(2\alpha^{\mathrm{int}}+2\alpha^{\mathrm{rot}}\right)+\gamma_2+\delta\epsilon_2,
\end{align}
where $\bm{\delta\epsilon}$ is an error on the ellipticity measurement. Equivalently to this, the components of shear after the rotation can be written as
\begin{align}
\centering
\epsilon_1^{\mathrm{withrot}}=&\epsilon^{\mathrm{int}}_1\cos(2\alpha^{\mathrm{rot}})-\epsilon^{\mathrm{int}}_2\sin(2\alpha^{\mathrm{rot}})\nonumber\\
\epsilon_1^{\mathrm{withrot}}=&\epsilon^{\mathrm{int}}_1\sin(2\alpha^{\mathrm{rot}})+\epsilon^{\mathrm{int}}_2\cos(2\alpha^{\mathrm{rot}})\text{.}
\end{align}
Note that, to first order in shape distortions, the order in which rotation and shear are applied to an image makes no difference.

The simplest idea would be to define a naive rotation estimator as the difference between the intrinsic and observed position angles
\begin{equation}
\label{eq:naive}
\hat{\alpha}^\text{rot}=\alpha^\text{obs}-\hat{\alpha}^\text{int}\text{,}
\end{equation}
with the observed position angle calculated in the standard fashion as $\alpha^\text{obs}=1/2\tan^{-1}(\epsilon^\text{obs}_2/\epsilon^\text{obs}_1)$. However, it is clear from equation (\ref{eq:obs_ellip2}) that the standard lensing effect of shear could induce an apparent rotation. Furthermore, the opposite effect also occurs: if a rotation signal is present and is not accounted for, then the estimate of the shear using the cBB estimator will be incorrect. Thus in this section, we will instead extend the cBB estimator to include the possibility of rotation. 

The $\chi^2$ as defined above (equation \ref{eq:chi2}) needs to be modified to include the rotation
\begin{align}\label{eq:chi2_2}
\chi^2=&\sum_{i=1}^Nw_i\biggl[\left(\epsilon_1^{\mathrm{obs},(i)}-\gamma_1\right)\sin\left(2\alpha_i^{\mathrm{int}}+2\alpha^{\mathrm{rot}}\right)\nonumber\\
&-\left(\epsilon_2^{\mathrm{obs},(i)}-\gamma_2\right)\cos\left(2\alpha_i^{\mathrm{int}}+2\alpha^{\mathrm{rot}}\right)\biggr]^2.
\end{align}
This in turn necessitates a change to the cBB estimator. Equations (\ref{eq:cBB})-(\ref{eq:h_matrix}) keep the same form, however the matrix $\mathbf{M}_i$ is now given by
\begin{equation}
\label{eq_modcbb}
\small
\mathbf{M}_i=\left(\hspace{-0.2cm}
\begin{array}{cc}
\beta_4^{\mathrm{int}} - \cos\left(4\hat{\alpha}_{i}^{\mathrm{int}}+4\alpha^{\mathrm{rot}}\right) & -\sin\left(4\hat{\alpha}_{i}^{\mathrm{int}}+4\alpha^{\mathrm{rot}}\right) \\
-\sin\left(4\hat{\alpha}_{i}^{\mathrm{int}}+4\alpha^{\mathrm{rot}}\right) & \beta_4^{\mathrm{int}} + \cos\left(4\hat{\alpha}_{i}^{\mathrm{int}}+4\alpha^{\mathrm{rot}}\right)
\end{array} \hspace{-0.2cm}\right)\text{.}
\normalsize
\end{equation}
This modified cBB estimator provides an estimate of the shear assuming that we know the rotation angle. 

It is possible to write down an estimator for the rotation by minimising the $\chi^2$ in equation (\ref{eq:chi2_2}) with respect to the rotation angle. However, this $\chi^2$ is identical upon making the substitution $\alpha^{\mathrm{rot}}\rightarrow\alpha^{\mathrm{rot}}+\pi$. Hence minimizing the $\chi^2$ with respect to $\alpha^{\mathrm{rot}}$ will lead to two degenerate estimates of the rotation angle $\hat{\alpha}^{\mathrm{rot}}$ and $\hat{\alpha}^{\mathrm{rot}}+\pi$. Assuming that the shear is small, we could use a naive rotation angle estimate (equation \ref{eqn:naive}) as a guide for which of these estimates to use.

A better method is to recover an estimator for $\alpha^{\mathrm{rot}}$ directly from equation (\ref{eq:obs_ellip2}) under the assumption that we know the shear. Rearranging this equation leads to
\begin{equation}\label{eq:obs_rear}
\alpha^{\mathrm{rot}}=\frac{1}{2}\tan^{-1}\left(\frac{\epsilon_2^{\mathrm{obs}}-\gamma_2}{\epsilon_1^{\mathrm{obs}}-\gamma_1}\right)-\alpha^{\mathrm{int}}.
\end{equation}
If we allow for random measurement errors on $\alpha^{\mathrm{int}}$ and $\bm{\epsilon}^{\mathrm{obs}}$ with the dispersion of the errors identical for the two components of the observed ellipticity, then the mean of the RHS of equation (\ref{eq:obs_rear}) for $N$ galaxies will provide an unbiased estimate of $\alpha^{\mathrm{rot}}$:
\begin{equation}\label{eq:rot_est}
\hat{\alpha}^{\mathrm{rot}}=\frac{1}{2}\tan^{-1}\left(\frac{\sum_{i=1}^N\sin\left(2\alpha_i^{\mathrm{new}}-2\hat{\alpha}_i^{\mathrm{int}}\right)}{\sum_{i=1}^N\cos\left(2\alpha_i^{\mathrm{new}}-2\hat{\alpha}_i^{\mathrm{int}}\right)}\right),
\end{equation}
where
\begin{equation}\label{eq:new}
\alpha_i^{\mathrm{new}}=\frac{1}{2}\tan^{-1}\left(\frac{\epsilon_2^{\mathrm{obs},(i)}-\gamma_2}{\epsilon_1^{\mathrm{obs},(i)}-\gamma_1}\right).
\end{equation}
We can now solve for $\bm{\gamma}$ and $\alpha^{\mathrm{rot}}$ simultaneously using the modification to the cBB estimator (equation (\ref{eq_modcbb})) and equation (\ref{eq:rot_est}). This combined estimator will be referred to as the full rotation estimator and is one of the main results of this paper. To our knowledge this comprises the first estimator for the rotation signal in cosmic-shear surveys.

\subsection{Properties of the estimators}
\label{sec_properties}
We now perform some simple numerical calculations in order to see how the estimators perform, including the cBB estimator and the two different rotation estimators:  the naive estimator (equation \ref{eq:naive}) and the full rotation estimator (equation \ref{eq:rot_est}). We simulate an area of sky from a survey by generating a number $N_\text{gal}$ of random galaxies with intrinsic ellipticities drawn randomly from a Gaussian distribution with a 1D dispersion of $0.3/\sqrt{2}$. We then choose input values for the shear and rotation and apply them to each galaxy in the simulation. We include an intrinsic alignment signal of $\gamma_1^{\mathrm{IA}}=0.02$ and $\gamma_2^{\mathrm{IA}}=-0.01$. For each galaxy, it is assumed that the intrinsic position angle of the galaxy is measured from polarisation data, with Gaussian errors on these measurements with standard deviation $\alpha_{\text{rms}}$. The different estimators are then applied to the galaxies in the simulated sky area, and this is repeated for $10^5$ simulated areas in order to determine the statistics of the estimators. For the results presented here, the parameters were fixed as $\gamma_1=-0.04$, $\gamma_2=0.03$, $\alpha_{\text{rms}}=10^{\circ}$ and $N_\text{gal}=100$, although varying these parameters did not affect the results. Three different values of the rotation signal in each sky area were used: $0^\circ$, $5^\circ$ and $30^\circ$. We also include a measurement error on the observed ellipticity, with the error drawn from a Gaussian with variance equal to the 1D intrinsic ellipticity dispersion.

\begin{figure}
\hspace{-0.7cm}\includegraphics[width=3.8in]{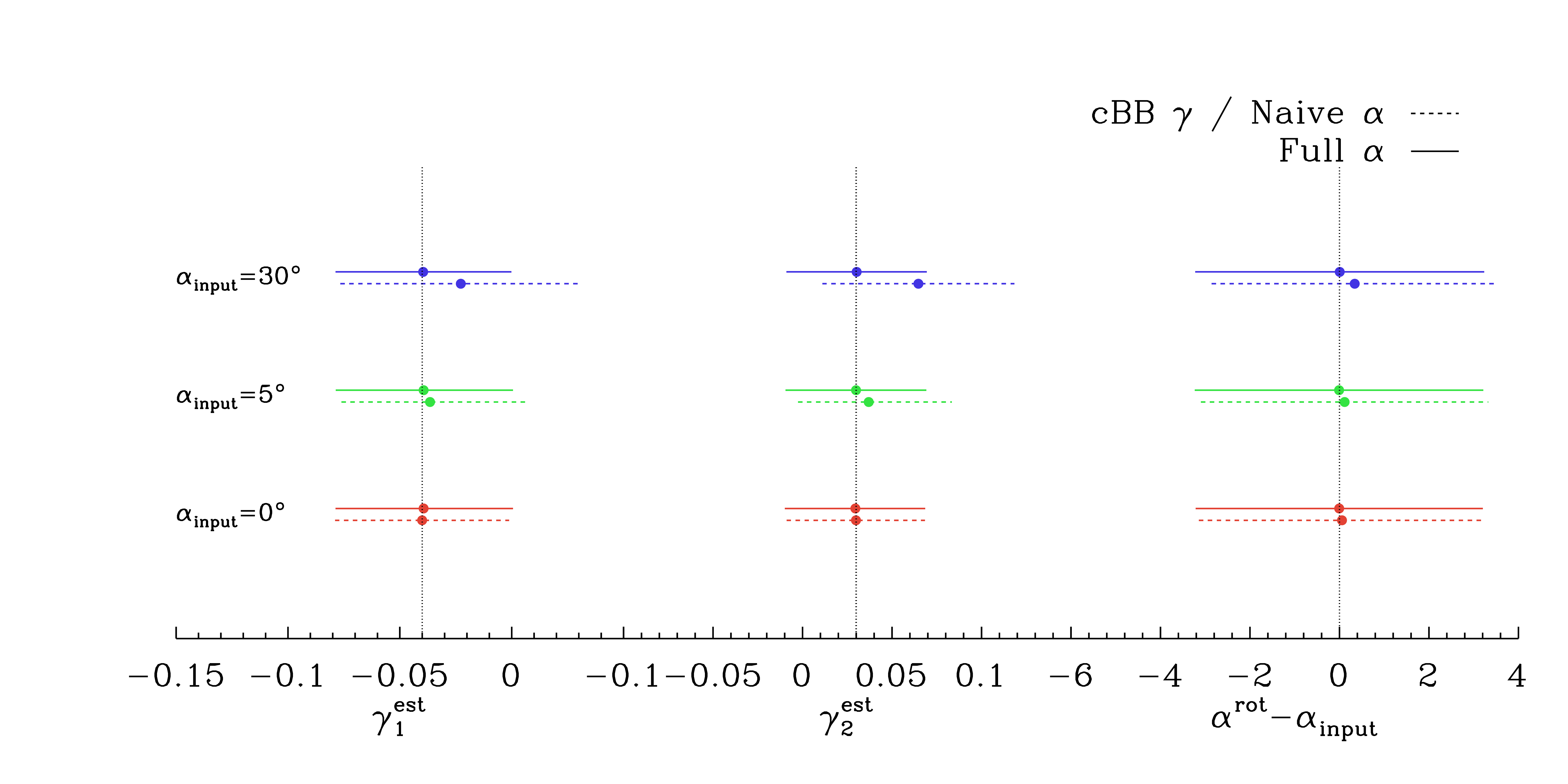}
\caption{The mean (central point) and dispersion (horizontal lines) of the distributions of the recovered estimates of the shear and rotation for three different values of the input rotation signal. The solid lines correspond to the full rotation estimator and the dashed lines correspond to the cBB estimator (for the shear) and the naive rotation estimator (for the rotation).}
\label{fig_simple_collated}
\end{figure}

In tables \ref{table_simplesims_bias} and \ref{table_simplesims_sigma} we present the results of these single pixel simulations for the three estimators for the three different values of the input rotation signal. Table \ref{table_simplesims_bias} shows the bias in the recovered estimates and table \ref{table_simplesims_sigma} shows the dispersion in the recovered estimates. The same information is presented graphically in figure \ref{fig_simple_collated}, where for each estimator and set of simulations the central point represents the mean of the recovered estimates and the line represents the dispersion.

The first result to notice from these simulations is that no information is lost by using the full rotation estimator: the dispersion of the shear estimates changes little when the full rotation estimator is used instead of the cBB estimator, even for the case of zero rotation signal. Furthermore, there is a bias present in both the shear (cBB) and naive rotation estimators when a rotation signal is indeed present.

The full rotation estimator performs well in these simulations. Not only does the rotation estimate have only a tiny bias,\footnote{There is a small expected residual bias due to the error on the $\alpha^{\mathrm{rot}}$ estimates propagating back into estimates of the shear. However, this bias is negligible in these tests, and it may be possible to correct for this in a similar way to the $\beta$ correction in the modified cBB estimator if we understand the error distribution on $\alpha^{\mathrm{rot}}$.} but the full rotation estimator also fixes the bias on the cBB shear estimates when a rotation is present. It is interesting to note that the dispersion of the cBB estimator increases as the rotation signal increases, this is not the case for the shear estimates from the full rotation estimator.

\begin{figure}
\includegraphics[width=2.5in]{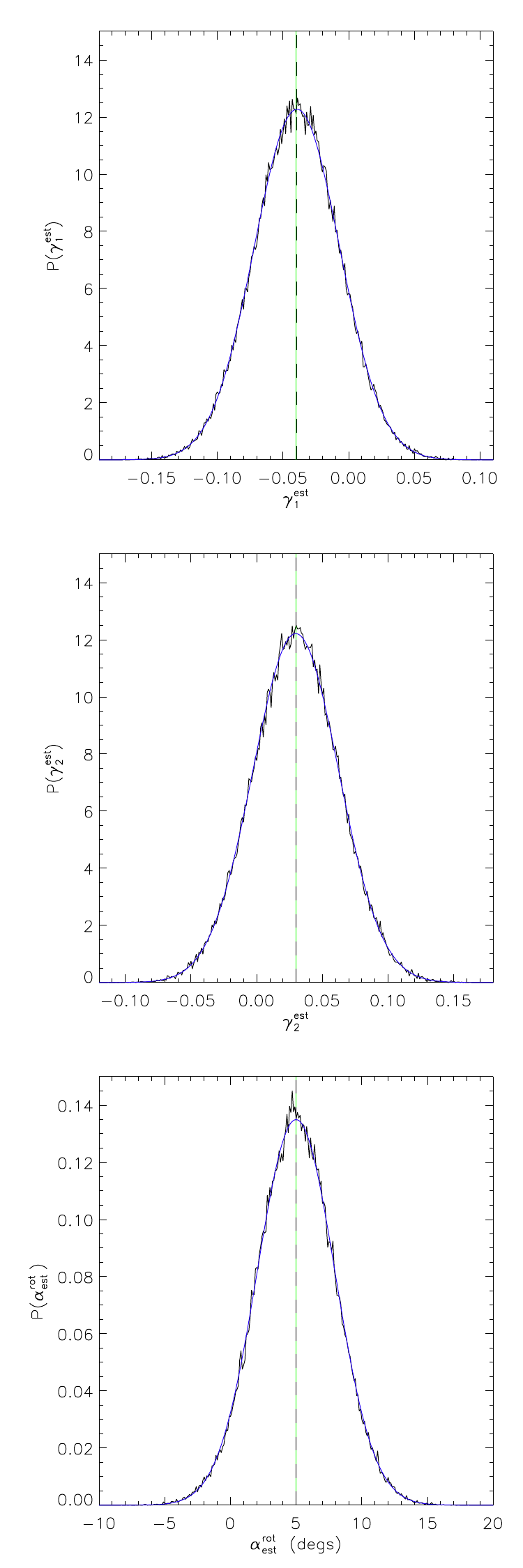}
\caption{The distributions of estimates recovered from $10^5$ realizations of a simulated sky area for the full rotation estimator with a $5^\circ$ rotation signal in each pixel. The black vertical line (dashed) shows the mean recovered estimate and the green line shows the input signal. The blue curves are best-fit Gaussian distributions for the recovered estimates, which fit the data well in all three cases.}
\label{fig_simple_newrot5}
\end{figure}

In figure \ref{fig_simple_newrot5} we show the distribution of estimates recovered from the $10^5$ pixel realisations using the full rotation estimator. In this figure, the dashed vertical (black) line shows the input value to the simulations, and the solid vertical (green) line shows the average from the estimator. The blue curve shows the best-fit Gaussian to the estimator. It can be seen from the plots that the dispersion of the estimator is well fit by a Gaussian distribution ( see appendix \ref{sec_errors} for a more detailed look at the dispersion of the estimator).

\begin{table*}
 \caption{Bias of the different estimators for the single pixel simulations in section \ref{sec_properties}. The input shear values are $\gamma_1=-0.04$, $\gamma_2=0.03$.}
\centering
\begin{mytabular}{|c||cc|cc|cc||} 
 \hline 
 \hline 
 &\multicolumn{2}{||c|}{Shear $\gamma_1$}& \multicolumn{2}{|c|}{Shear $\gamma_2$} &  \multicolumn{2}{|c||}{Rotation $w$}  \\
Rotation signal &   cBB & full rotation &  cBB & full rotation  &  naive & full rotation    \\
\hline
$0 ^\circ$ & $-3.75\times10^{-5}$ & $6.18\times10^{-4}$ & $-1.51\times10^{-4}$ & $-4.84\times10^{-4}$ & $5.64\times10^{-2}$ & $-6.14\times10^{-3}$ \\ 
$5 ^\circ$ & $3.50\times10^{-3}$ & $6.25\times10^{-4}$ & $6.97\times10^{-3}$ & $-1.11\times10^{-4}$ & $1.15\times10^{-1}$ & $-8.76\times10^{-3}$\\
$30 ^\circ$ & $1.72\times10^{-2}$ & $4.25\times10^{-4}$ & $3.47\times10^{-2}$ & $2.38\times10^{-4}$ & $3.41\times10^{-1}$ & $5.79\times10^{-3}$\\
 \hline 
 \end{mytabular} 
\label{table_simplesims_bias}
\normalsize
\end{table*}

\begin{table*}
 \caption{Dispersion of the different estimators for the single pixel simulations in section \ref{sec_properties}. The input shear values are $\gamma_1=-0.04$, $\gamma_2=0.03$.}
\centering
\begin{mytabular}{|c||cc|cc|cc||} 
 \hline 
 \hline 
 &\multicolumn{2}{||c|}{Shear $\gamma_1$}& \multicolumn{2}{|c|}{Shear $\gamma_2$} &  \multicolumn{2}{|c||}{Rotation $w$}  \\
Rotation signal &   cBB & full rotation &  cBB & full rotation  &  naive & full rotation    \\
\hline
$0 ^\circ$ & $3.90\times10^{-2}$ & $3.94\times10^{-2}$ & $3.88\times10^{-2}$ & $3.95\times10^{-2}$ & $3.20$ & $3.21$ \\ 
$5 ^\circ$ & $3.96\times10^{-2}$ & $3.93\times10^{-2}$ & $3.95\times10^{-2}$ & $3.94\times10^{-2}$ & $3.21$ & $3.23$\\
$30 ^\circ$ & $5.39\times10^{-2}$ & $3.95\times10^{-2}$ & $5.37\times10^{-2}$ & $3.92\times10^{-2}$ & $3.20$ & $3.23$\\
 \hline 
 \end{mytabular} 
\label{table_simplesims_sigma}
\normalsize
\end{table*}

\section{Cosmological simulations}
\label{sec_cosmosims}
In addition to the single pixel simulations of the previous section, we have performed more detailed cosmological simulations of weak lensing skies in order to test the estimator in a more realistic setting. The simulations in this section are full sky simulations, similar to those run in \cite{brown2011} and \cite{whittaker2015}, see there for the full details. We will briefly present some of the relevant details here; note that we do not model any systematics from the instrument.

Firstly, Healpix \citep{gorski2005} is used to split the sky into pixels with a pixel size of $3.4 \text{arcmin}$ (Healpix N$_\text{side}$ parameter is 1024). We consider three tomographic redshift bins, with 46 galaxies per pixel per redshift bin. This corresponds to a total of approximately $12$ galaxies per square arc minute, summed over all three redshift bins. The redshifts of the bins are as follow: $0 < z_1 < 1.58941$, $1.58941 < z_2 < 2.46108$ and $2.46108 < z_3$, with the redshifts of sources assigned a random error with $\sigma_z=0.05(1+z)$. The source distribution follows \cite{smail1994} , with $\beta=1.5$ and $\bar{z}=1.42$.

The rotation and shear signals are taken to be constant in each pixel, and are generated from Gaussian random fields based on theoretical power spectra using Healpix. The power spectra used as inputs consist of a $\Lambda$CDM $E$-mode shear spectrum and a cosmic string $B$-mode/rotation spectrum (see section \ref{sec_strings}). Note that, up to large-scale relativistic corrections \citep{bernardeau2010}, the $B$-mode and rotation power spectra are equal in the same way that the convergence and $E$-mode spectra are equal. The $\Lambda$CDM parameters used correspond to the \emph{Planck} best fit cosmology (see the final column of table 4 in \cite{planckparams2015}): $\omega_b=0.0223$, $\omega_c=0.1188$, $\ln 10^{10}A_s=3.064$, $n_s=0.9667$, $\tau=0.066$, $H_0=67.74$ and a flat universe is assumed. The transfer function fitting formulae from \cite{eisenstein1999} are used to generate the E-mode spectra and halofit \citep{smith2003} is used for non-linear corrections to these power spectra. We will also consider intrinsic alignment fields, using the non-linear alignment model \citep{bridle2007} normalised to the SUPERCOSMOS level, following \cite{brown2011}\footnote{We note that the non-linear alignment model only sources $E$-modes of cosmic shear. For simplicity and for comparison to previous work, we will use this model here, particularly as it is amongst the most typically used model of intrinsic alignments. However, we note that there do exist models of intrinsic alignments where a $B$-mode of cosmic shear is generated, see e.g. \cite{crittenden2001}.}. The contribution of the cosmic strings to the $E$-mode spectrum is much smaller than the $\Lambda$CDM spectrum and is ignored here for simplicity.

Once the shear (including both $E$- and $B$-mode components) and rotation fields in each pixel have been calculated, the code generates the appropriate number of galaxies in each pixel with random intrinsic ellipticity, and an estimate is assumed of the intrinsic position angle of each galaxy from polarisation measurements. The error on the estimate of the intrinsic position angle is taken to be Gaussian with standard deviation $\alpha_\text{rms}=5^\circ$, this is increased to $\alpha_\text{rms}=10^\circ$ for some of the runs. If the run involves intrinsic alignments, then these are added at this stage too. The shear and rotation signals in that pixel are then applied to each galaxy (note that to linear order in shape distortions it does not matter which order these are applied in). Thus, there are two primary sources of noise: the intrinsic shapes of galaxies and the scatter in the polarisation to intrinsic position angle relation.

The different shear and rotation estimators are then applied to the galaxies in each pixel that has been generated, and the power spectra of the reconstructed fields are then calculated using Healpix. The simulations are only analysed up to $\ell_\text{max}=2048$ in order to avoid the strongly non-linear regime. As in \cite{brown2011} and \cite{whittaker2015}, the estimators have a noise bias contribution that is removed using simulations, however note that we did not apply this correction in several of the simulations in order to illustrate one of the advantages of using the rotation estimator.

The full rotation estimator is solved numerically using an implementation of Broyden's method \citep{broyden1965}. We will present results for the cBB estimator, the traditional cosmic shear estimator that averages over the ellipticity in a pixel, and our full rotation estimator.

\subsection{Cosmic string spectra}
\label{sec_strings}
In order to demonstrate the use of the rotation estimator, we require a source for the rotation signal. Cosmic strings are known to produce vector perturbations to the metric, so here we will briefly present details of the cosmic string spectra that were used for our simulations.

Cosmic strings, first predicted in the context of symmetry breaking phase transitions in the early Universe \citep{kibble1976}, arise as topological defects along lines where a complex field has remained trapped in a false vacuum after a symmetry breaking phase transition. Before the first acoustics peaks of the CMB were detected \citep{netterfield1996,mauskopf2000}, cosmic strings provided an `active' alternative to the `passive' structure formation scenario based on inflationary generated perturbations. Nowadays, a subdominant contribution from cosmic strings is still viable \citep[see e.g.][]{battye2006}. Moreover, a renewed interest in cosmic strings has also been driven by the possibility that many string theory models predict the generation of macroscopic strings at the end of `stringy' inflation \citep{copeland2003,dvali2004}.

Note that a cosmic string network will in general source both vector and tensor perturbations. We focus here on the vector perturbations, which are thought to be approximately an order of magnitude larger than their tensor counterparts \citep{contaldi1998}. For our spectra, we follow the so-called velocity-dependent one-scale model \citep[e.g.][]{martins1996,martins2000}. In this model a cosmic string network is characterised by a correlation length, $\xi=1/(H\gamma_s)$ (where $H$ here is the Hubble factor), and a root-mean square velocity, $v_{\rm rms}$. In the limit where the cosmic string network approaches a scaling solution, $\gamma_s$ and $v_{\rm rms}$ can be assumed to be constant.

The use of cosmic shear $B$-modes or rotation to constrain cosmic strings was first suggested in \cite{thomas2009}. Since then, the use of the $B$-modes to constrain cosmic strings has been more fully explored in \cite{yamauchi2013} and \cite{yamauchi2012} using more detailed models for the cosmic string metric perturbations. Here, we will use the calculation of the cosmic string shear $B$-modes from \cite{yamauchi2013}. As we are ignoring tensor perturbations, the $B$-modes are the same when calculated with either the geodesic deviation equation or displacement approximation approach, and this is borne out explicitly for the vector perturbations in \cite{yamauchi2013} and \cite{yamauchi2012}. As shown above, the power spectrum of the rotation mode equals that of the $B$-mode when using the displacement approximation. Thus, we will use the $B$-mode spectra presented here as the input to our simulations for both the $B$-mode and rotation mode spectra.

Here we will just summarise the relevant results from  \cite{yamauchi2013}, see that paper for the full details. Following \citet{yamauchi2013}, the contribution to  the cosmic shear $B$-modes from the vector perturbations can be written as a sum over the two spins ($m=\pm1$) of the vector perturbations
\begin{eqnarray}\label{eq:C_l^B}
&C_\ell^B=4\pi\int\!\!\de\ln k\,\sum_{m=-1,1}\mathcal M_{\ell m}^B(k)\\
&\mathcal M_{\ell m}^B(k)=\int\!\!\de\chi_1\de\chi_2\,k^2\mathcal S_{\ell m}^X(k,\chi_1)\mathcal S_{\ell m}^B(k,\chi_2)\Delta^2_{|m|}(k;\chi_1,\chi_2)\text{.}\nonumber
\end{eqnarray}
Here, $\Delta^2_{|m|}$ is the dimensionless power spectrum of cosmic string vector perturbations, which is the same for the two values of the spin. The $B$-mode transfer function $\mathcal S_{\ell m}^B$ is given by
\begin{equation}\label{eq:S_vec}
\mathcal S_{\ell,\pm1}^B=-\sqrt{\frac{(\ell+2)(\ell-1)}{2}}\frac{1}{k\chi}\int_\chi^\infty\!\!\de\chi'\,\frac{\de N_g}{\de\chi'}\left(\pm\frac{1}{2}\right)j_\ell(k\chi')\text{,}
\end{equation}
where $j_\ell(k\chi)$ is the spherical Bessel function and $\de N_g(\chi)/\de\chi=\de N_g(z)/\de z$ is the source redshift distribution, normalised such that its integral over the entire redshift range equals unity.

The dimensionless power spectrum of cosmic string vector perturbations is given by
\begin{equation}
\Delta^2_1(k,\chi)=\frac{256\pi\sqrt{6\pi}}{3}(G\mu)^2\frac{v_{\rm rms}^2}{1-v_{\rm rms}^2}\frac{\chi^2a^9(\chi)}{k^2H\xi^5}\mathrm{erf}\left[\frac{k\xi}{2\sqrt{6}a(\chi)}\right]\text{,}
\end{equation}
where $G\mu$ is the dimensionless string tension and $\text{erf}(x)=\frac{2}{\sqrt{\pi}}\int^x_0 dy e^{-y^2}$ is the error function. We use the approximate expressions
\begin{align}
\gamma_s&\approx\sqrt{\frac{\sqrt{2}\pi}{3\tilde c\mathcal P}},\\
v_{\rm rms}^2&\approx\frac{1}{2}\left(1-\frac{\pi}{3\gamma_s}\right),
\end{align}
where $\tilde c\approx0.23$ is the efficiency of the loop information and $\mathcal P$ is the inter-commuting probability \citep[see also][\S~5]{namikawa2011}, for which we used the value $\mathcal P=10^{-3}$.

\begin{figure}
\includegraphics[width=2.5in, angle=270]{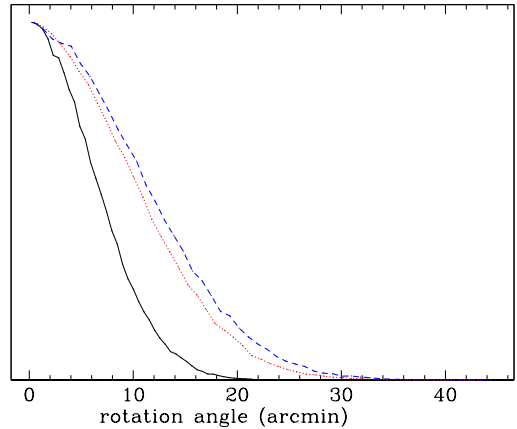}
\caption{Probability distribution of rotation values in pixels for cosmic string spectra with $G\mu=1.3\times10^{-7}$. The rotation angles are in arc minutes and the three curves correspond to the three redshift bins: the solid (black) curve is the lowest redshift bin, the dotted (red) curve is the middle redshift bin and the dashed (blue) curve is the highest redshift bin. For these bins, the mean rotation angles are approximately $4.9$ arcmin for the first redshift bin, $7.5$ arcmin for the second redshift bin and $8.2$ arcmin for the third redshift bin.}
\label{fig_rot_angles}
\end{figure}

The cosmic string spectra used as the input for the $B$-mode and rotation signals scale with the amplitude of the string tension $G\mu$, such that $\Delta^2_1(k,\chi)\propto(G\mu)^2$. For our simulations we use two values of $G\mu$: we use an illustrative value $G\mu=10^{-6}$ for exploring the properties of the estimator further and investigating biases. We also investigate the more realistic value $G\mu=1.3\times10^{-7}$. Figure \ref{fig_rot_angles} shows the distribution of rotation angles in pixels for the spectrum with $G\mu=1.3\times10^{-7}$, where the rotation angles are measured in arc minutes. The value $G\mu=1.3\times10^{-7}$ is chosen in line with the strongest constraints from \emph{Planck} \citep{planckstrings}. This constraint depends on the details of the cosmic string model and more than one model is considered in the \emph{Planck} papers \cite{planckstrings}. More precisely, the \emph{Planck} constraints range from $1.3\times10^{-7}-10.5\times10^{-7}$ at the 2$\sigma$ level. Thus, our two values of $G\mu$ capture the range of the \emph{Planck} constraints well.

\section{Results}
\label{sec_results}
We now present the results from the cosmological simulations. For brevity we only show the spectra for one combination of $G\mu$ and $\alpha_\text{rms}$, however we present several plots containing the bias and errors of the different estimators for different simulation parameters.

\begin{figure*}
\includegraphics[width=7in]{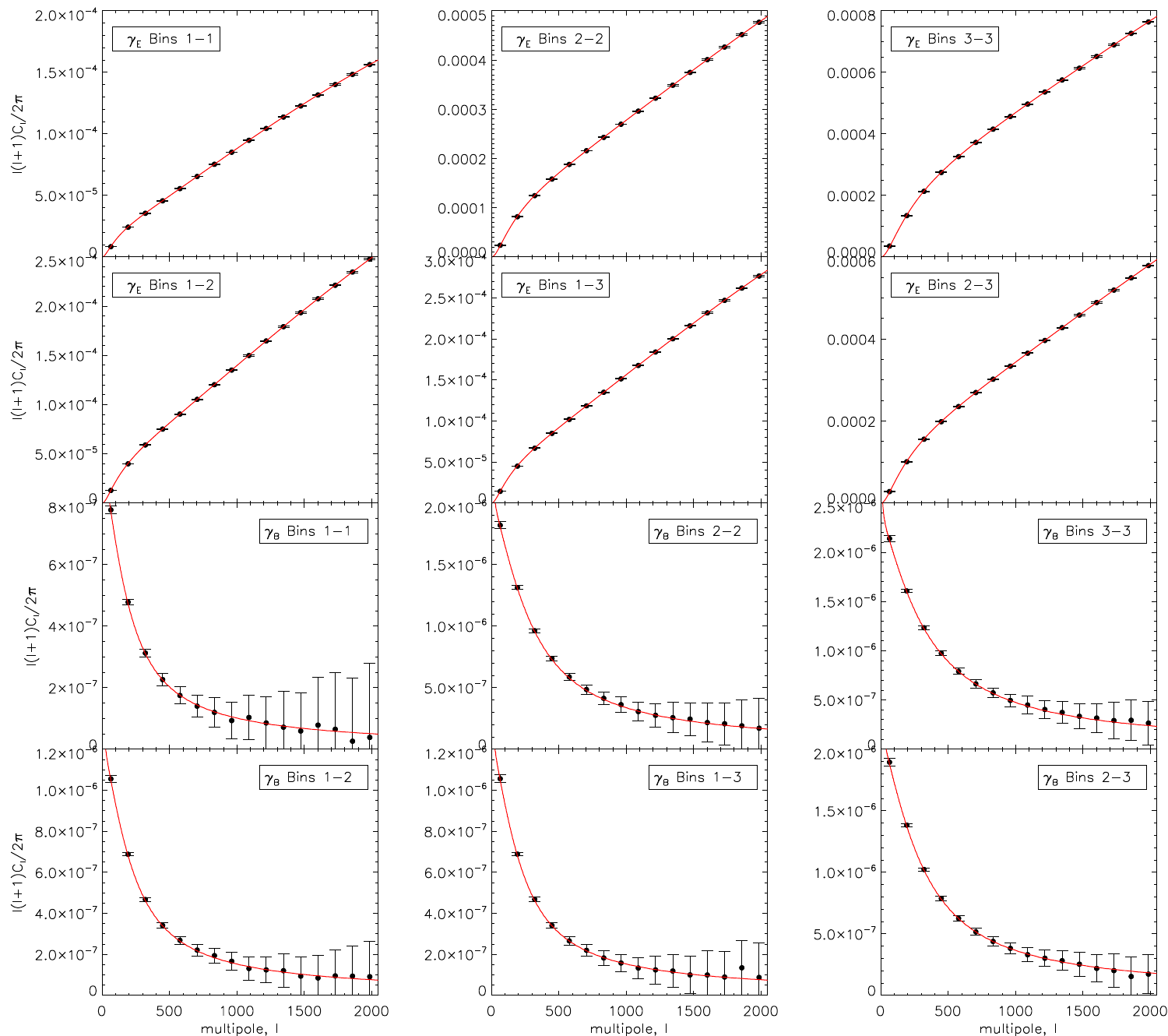}
\caption{The $E$- and $B$-modes of shear recovered using the full rotation estimator when intrinsic alignments are present. The cosmic string spectra have amplitude $G\mu=1.3\times10^{-7}$ and the error on estimates of the intrinsic position angle is $\alpha_\text{rms}=5^\circ$. The first two rows correspond to the $E$-modes and the bottom two rows correspond to the $B$-modes, with the first and third rows showing the auto-correlation spectra in the three different redshift bins and the second and fourth rows showing the cross-correlation spectra between different redshift bins.}
\label{fig_cosmo_newgmu13_ia}
\end{figure*}

\begin{figure*}
\includegraphics[width=7in]{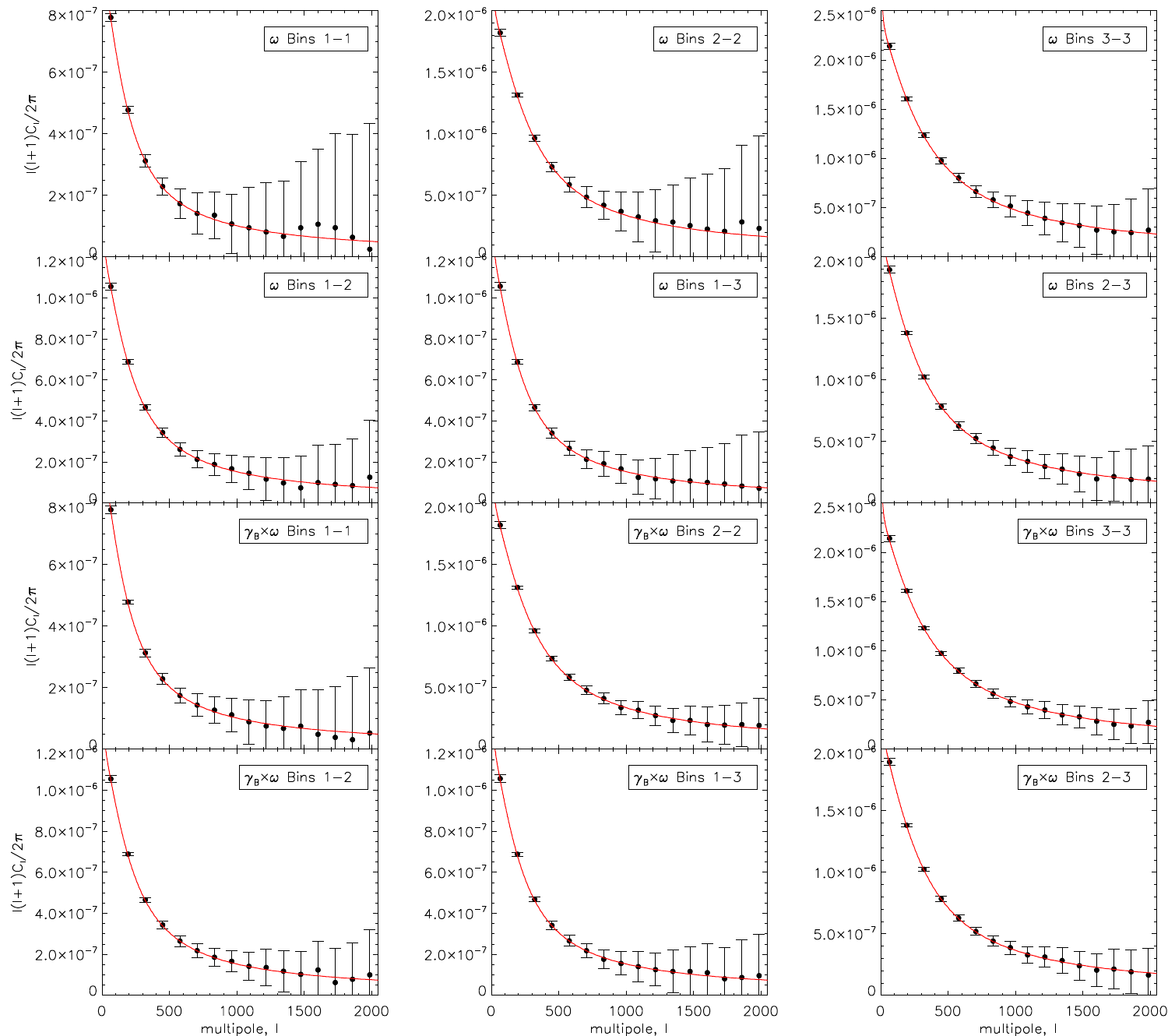}
\caption{The rotation spectrum and rotation-$B$-mode cross correlation recovered using the full rotation estimator when intrinsic alignments are present. The cosmic string spectra have amplitude $G\mu=1.3\times10^{-7}$ and the error on estimates of the intrinsic position angle is $\alpha_\text{rms}=5^\circ$. The first two rows correspond to the rotation and the bottom two rows correspond to the rotation $B$-mode cross correlation, with the first and third rows showing the auto-correlation spectra in the three different redshift bins and the second and fourth rows showing the cross-correlation spectra between different redshift bins.}
\label{fig_cosmo_newgmu13_wb_ia}
\end{figure*}

In figures \ref{fig_cosmo_newgmu13_ia} and \ref{fig_cosmo_newgmu13_wb_ia} we show the power spectra recovered by the full rotation estimator for the case $G\mu=1.3\times10^{-7}$ and $\alpha_\text{rms}=5^\circ$ where intrinsic alignments are included. In figure \ref{fig_cosmo_newgmu13_ia} we show the recovered $E$- and $B$-mode shear spectra across the three redshift bins and their cross correlations and figure \ref{fig_cosmo_newgmu13_wb_ia} shows the recovered rotation and rotation $B$-mode cross correlation spectra. These plots show that the full rotation estimator recovers the $E$-mode of shear for a realistic cosmological signal. In addition, the $B$-mode and rotation spectra, and their cross spectrum, can also be detected across a significant fraction of the $l$-range for most of the redshift bins. The detection of these signals is better on larger scales and at higher redshifts, as the signal is larger in these regimes. The errors on the $B$-mode, rotation and cross spectra are similar, although slightly worse for the rotation.

\begin{figure}
\includegraphics[width=3in]{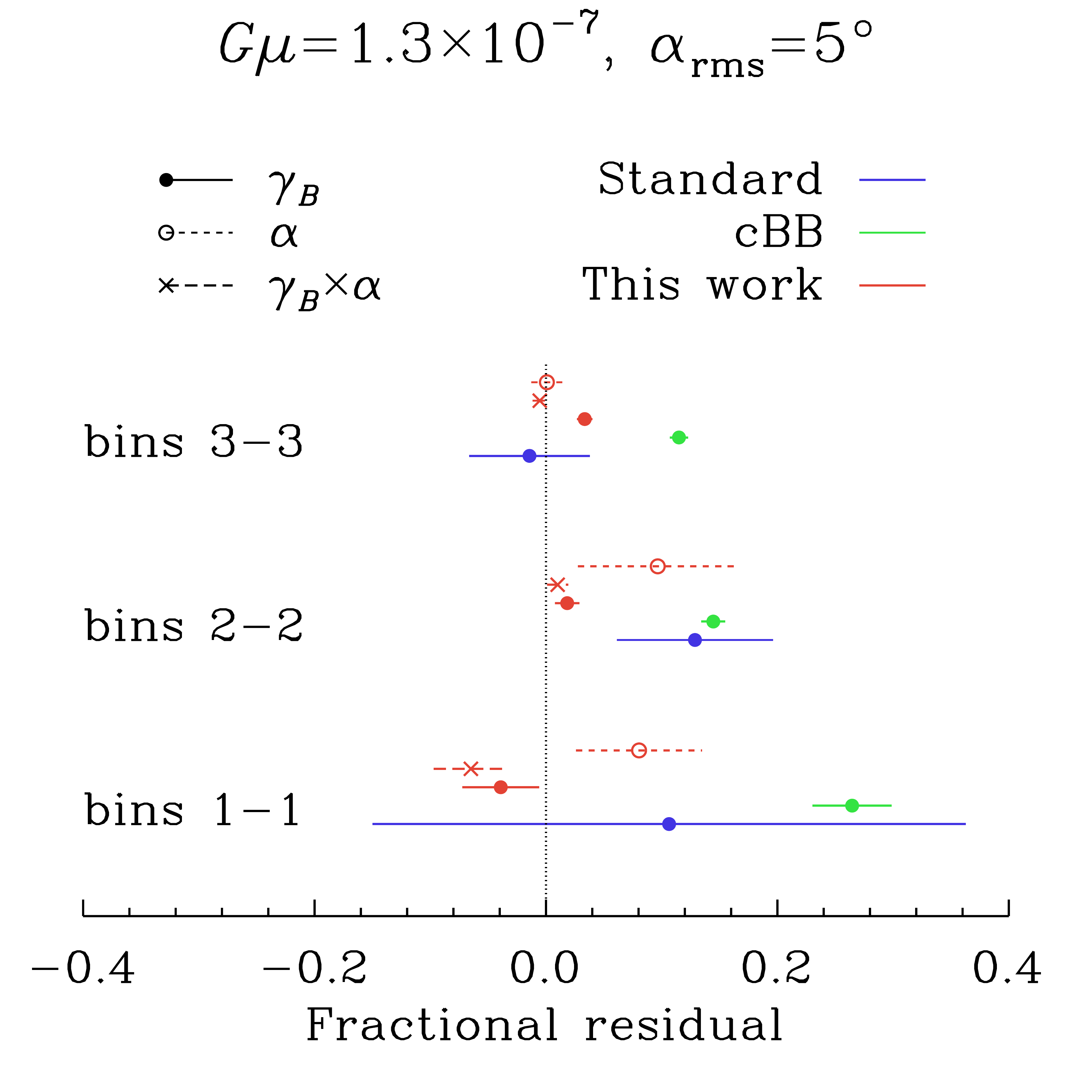}
\caption{The bias (shown by the location of the central point) and the errors (shown by the horizontal line) for different estimators when the cosmic string spectra have amplitude $G\mu=1.3\times10^{-7}$ and the error on estimates of the intrinsic position angle is $\alpha_\text{rms}=5^\circ$. The solid circles correspond to the $B$-mode spectra for the standard shear estimator (blue), cBB estimator (green) and full rotation estimator (red) The additional red points correspond to the full rotation estimator rotation spectrum (open circles) and $B$-mode-rotation cross spectrum (crosses). The three rows correspond to the auto-correlation spectra for the different redshift bins.}
\label{fig_spectra_collate1}
\end{figure}

In order to compare the different estimators, in figure \ref{fig_spectra_collate1} we present the bias and error bars on the recovered spectra averaged over the $l$ bins. We note that this is a significant data compression step compared to figures \ref{fig_cosmo_newgmu13_ia} and \ref{fig_cosmo_newgmu13_wb_ia}, however it allows for easy comparison of the different estimators. In this figure we present the bias and error for the $B$-mode spectrum for the standard shear estimator, cBB estimator and full rotation estimator. In addition, we present the rotation and rotation B-mode cross spectra for the full rotation estimator. We have again chosen the case $G\mu=1.3\times10^{-7}$ and $\alpha_\text{rms}=5^\circ$ and we show the auto correlation spectra for the three redshift bins. As expected from the single pixel simulations, we can see that the cBB estimator is clearly biased, but has a similar error to the $B$-mode from the full rotation estimator. The standard estimator has a significantly higher error on the $B$-mode. This plot also shows the small residual bias on the full rotation estimator and that the errors on the rotation are larger than those on the $B$-mode, as mentioned above. Note that we don't show the $E$-mode spectra here. The errors on the $E$-mode spectra are small for all estimators, however our simulations demonstrate the known bias to the standard estimator due to the intrinsic alignments. As shown in \cite{brown2011,whittaker2015}, the cBB estimator removes this bias. Our simulations show that the full rotation estimator preserves this very useful feature of the cBB estimator.

\begin{figure}
\includegraphics[width=3in]{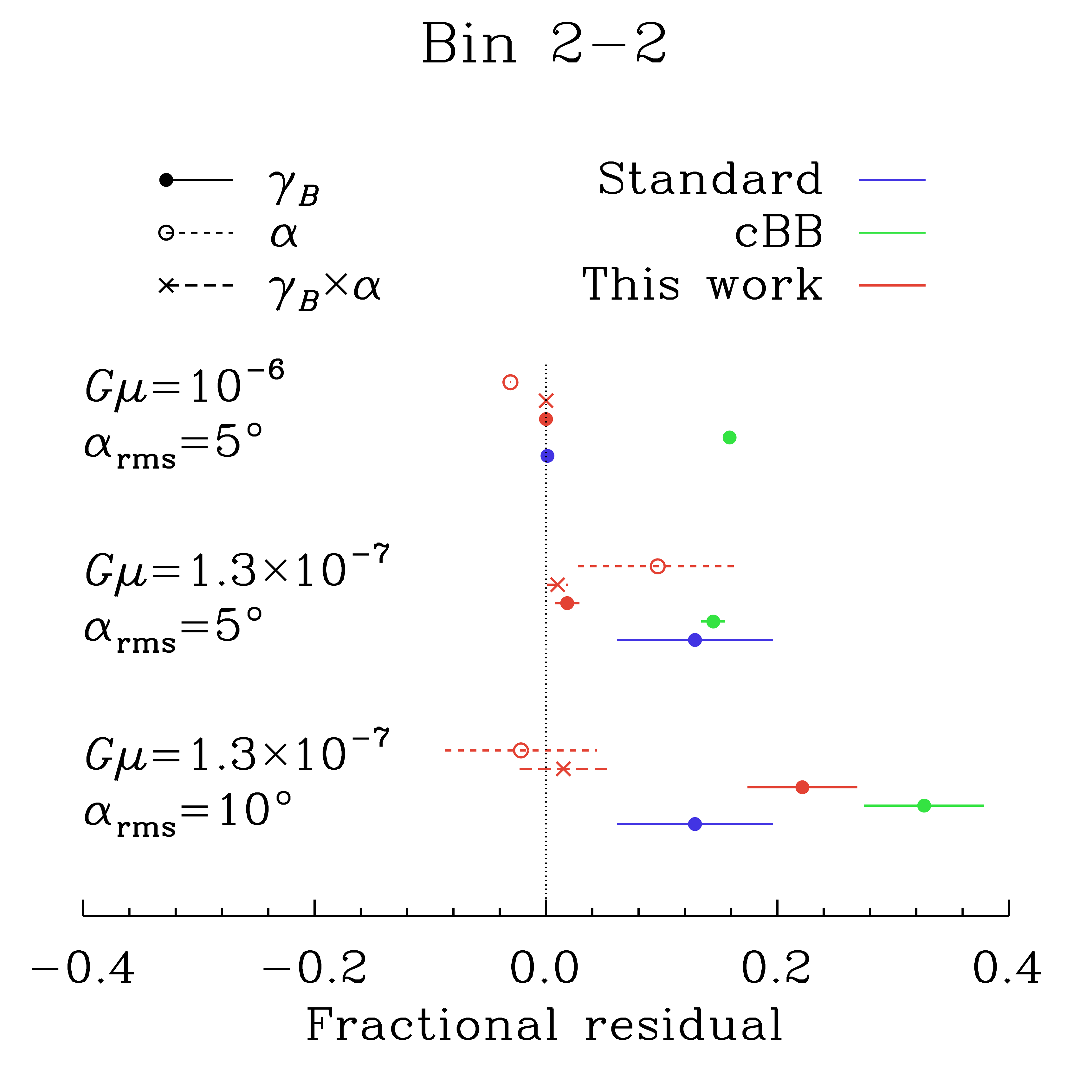}
\caption{The bias (shown by the location of the central point) and the errors (shown by the horizontal line) for different estimators for the auto-correlation spectrum for the second redshift bin. The solid circles correspond to the $B$-mode spectra for the standard shear estimator (blue), cBB estimator (green) and full rotation estimator (red) The additional red points correspond to the full rotation estimator rotation spectrum (open circles) and $B$-mode-rotation cross spectrum (crosses). The three rows correspond to three different choices of $G\mu$ and $\alpha_\text{rms}$.}
\label{fig_spectra_collate2}
\end{figure}

In figure \ref{fig_spectra_collate2} we examine how the bias and errors change for different values of $G\mu$ and $\alpha_\text{rms}$. In this figure, we display only the auto-correlation spectra for the second redshift bin for simplicity. As expected, the errors on the cBB and full rotation estimators worsen as $\alpha_\text{rms}$ increases. The relative errors for all estimators are much smaller for $G\mu=10^{-6}$, also as expected. A spectrum of this amplitude is ruled out for some of the cosmic string models considered in \cite{planckstrings}, but is still allowed for others. There is a clear bias on the cBB estimator in all cases.

In most of these simulations, we subtracted the noise bias following \cite{brown2011,whittaker2015}. However, one of the advantages of cross correlations is that this procedure is not required. We have verified that this applies to the rotation $B$-mode cross correlation using simulations run without the noise bias correction. Thus, these spectra could act as a crucial check of the noise bias subtraction methodology in future surveys.  We also used our simulations to check the result from the single pixel simulations that the errors on the shear do not worsen compared to the cBB estimator when the full rotation estimator is used. This result continues to hold, even for the case where the input rotation signal is zero.

We have run the simulations without a contribution from intrinsic alignments and find little difference for either the cBB or full rotation estimators. As mentioned above, we recover the result that the standard shear estimator is biased in the presence of intrinsic alignments.

\section{Conclusion}
\label{sec_conc}
In this paper, we have presented an estimator for the rotation mode in radio weak-lensing surveys, equation \ref{eq:rot_est}. This is the first time that an estimator for the rotation signal in cosmic-shear surveys has been proposed. This work builds on the proposal of \cite{brown2011} to use information on the polarisation of radio sources to determine the intrinsic, un-lensed, position angle of radio galaxies. As the rotation mode has not yet been constrained, this estimator provides a null-test of $\Lambda$CDM and the assumptions that underly weak-lensing surveys.

We have applied this estimator on simulated weak-lensing skies, showing that the cBB estimator for the shear \citep{whittaker2015} is biased in the presence of rotation and also that the full estimator we have proposed costs almost nothing in terms of information on the shear, even when there is no rotation present. Furthermore, using cosmic strings as an example source for the rotation mode, we have shown that this estimator could detect a cosmic string network that is compatible with the constraints from the \emph{Planck} satellite. We have shown how the errors change for different values of the string amplitude $G\mu$ and for different values of the intrinsic scatter in the polarisation angle-intrinsic position angle relationship. Although we focussed on cosmic strings as a pedagogical example here, we also note that the presence of bi-refringence would create an effective rotation signal, due to its effect on the relationship between the galaxy polarisation and the intrinsic position angle of the galaxy. Thus, our estimator will also act as a null test for the presence of bi-refringence.

The relationship between the $B$-mode and the rotation means that recovering both spectra allows for a crucial consistency check of cosmic-shear surveys, that could highlight systematics. This is improved by examining the cross spectrum between the $B$-mode and the rotation, which is also recovered by our full rotation estimator. In addition, we found that this cross correlation does not require noise bias to be subtracted, creating a crucial check of the subtraction of the noise bias from the auto-correlation spectra. Although not done here, we note that the rotation and $B$-mode measurements could be combined in order to yield stronger constraints on beyond $\Lambda$CDM physics. 

We found that the full rotation estimator preserves the property of the cBB estimator that the recovered spectrum is not biased by intrinsic alignments, in contrast to the standard shear estimator. However, we note that the intrinsic alignment model used here consists of an $E$-mode signal only. It is beyond the scope of this paper to investigate in detail the different effects of different models for the intrinsic alignments, including those that may generate $B$-modes.

Even assuming that we live in a $\Lambda$CDM universe, there is a cosmological source of the $B$-mode and rotation signal. This is the second order effect of the Born correction/lens-lens coupling terms in the deflection angle \citep{krause2010}. However, although this signal scales differently with $l$ compared to the cosmic string signal considered here, it is below the level of this cosmic string signal on all scales and is unlikely to be detectable using currently planned radio surveys. Nonetheless, as mentioned above, the estimator that we have proposed provides some useful consistency checks for weak-lensing surveys even in the absence of a rotation signal.

Finally, we note that there has been much recent interest in relativistic effects on large scales in cosmological observables. As shown in \cite{bernardeau2010}, a difference between the $E$-mode and convergence is created on large scales. We expect that a similar difference might appear between the $B$-mode and the rotation signal on such scales. This difference has not yet been calculated, but it is possible that this difference between the rotation and the $B$-mode could be detectable with future large scale radio surveys.

{\sl Acknowledgements} DBT is supported by the European Research Council under the European UnionÕs
Seventh Framework Programme (FP7/2007-2013) / ERC
Grant Agreement n. 617656 ÒTheories and Models of the
Dark Sector: Dark Matter, Dark Energy and GravityÓ. DBT thanks David Bacon for encouragement during the early stages of this work. LW, SC and MLB are supported by an ERC Starting Grant (grant no. 280127). MLB also acknowledges the support of a STFC Advanced/Halliday fellowship (grant number ST/I005129/1).

\appendix
\section{Rotation of polarisation angle by vectors}
\label{app_faraoni}
We follow \cite{faraoni2008}, but allow for arbitrary first order metric perturbations. The photon wavevector $S^{\mu}$ and polarisation vector $P^{\mu}$ have the same form as in \cite{faraoni2008} at zeroth order, $S^{\mu}=(1,0,0,1)=\nabla^{\mu} S$ and $P^{\mu}=(0,1,0,0)$ respectively. Here, $S$ is the phase of the photon and the amplitude of the photon is expressed as $\hat{A}_{\mu}=a P^{\mu}$, with $a$ complex and $P^{\mu}$ real. The equation that is integrated in \cite{faraoni2008} is 
\begin{equation}
\label{eqn_faraoni}
\frac{d P^{\mu}}{d\sigma}=\frac{1}{2}\left(\frac{P^{\nu}\partial_{\nu}a}{a} +\nabla_{\nu}P^{\nu}\right) S^{\mu}\text{,}
\end{equation}
where $\sigma$ is the affine parameter. The background value of $a$ is $A/\sigma$, where $A$ is a complex constant. Due to the zeroth order form of $P^{\mu}$, the term in brackets has no zeroth order term, so the only contributing terms involve the zeroth order $S^{\mu}$, and hence require $\mu=0,3$. Thus, integrating equation \ref{eqn_faraoni} for the spatial directions transverse to the line of sight ($\mu=1,2$) will result in just a constant of integration, and the only effect on the photon polarisation is thus a boundary term. This result is independent of which types of perturbations are in the metric, so it includes vector perturbations.

\section{Errors on the estimates}\label{sec_errors}
In this appendix we show an approximate iterative method to calculate the error on the rotation estimator discussed in the main text. We will also describe a simplification of the method and investigate the accuracy of the method with and without the simplification. The equations are initially presented without derivation in order to improve the clarity of the algorithm. For the interested reader, the derivation of the equations is presented in \ref{app_deriv}.
\subsection{Calculating the error on the estimator}
The rotation estimator is coupled to the shear estimator, which results in their errors being coupled as well. The error on the shear can be approximated as 
\begin{equation}\label{eq:err_cbb_dt2}
\sigma_{\hat{\gamma}}^2=\frac{1}{N\beta_4^{\mathrm{int}^2}}\left\{\left[1+\beta_4^{\mathrm{int}^2}\left(1-2\beta_4^{\mathrm{rot}}\right)\right]\sigma_{\epsilon}^2+\left(1+\beta_4^{\mathrm{int}^2}\right)\sigma^2\right\},
\end{equation}
where $\sigma$ is the measurement error on the galaxy shapes and $\sigma_{\epsilon}$ is the 1D dispersion of the intrinsic ellipticities of the sources. As in the main text,  $N$ is the number of galaxies in each cell on the sky in which the shear is estimated. Under the assumption that the errors on $\alpha^{\mathrm{int}}$ (i.e. the errors on the intrinsic position angles from the polarisation data) are Gaussian distributed, the $\beta_n^{\mathrm{int}}$ functions are given by
\begin{equation}\label{eq:beta_gauss}
\beta_n^{\mathrm{int}}=\exp\left(-\frac{n^2}{2}\alpha_{\mathrm{rms}}^2\right),
\end{equation}
where $\alpha_{\mathrm{rms}}$ is the standard deviation of the Gaussian errors. Similarly, under the assumption that the distribution of errors on $\alpha^{\text{rot}}$ ($f_{\delta\alpha^{\mathrm{rot}}}$) is Gaussian,
\begin{equation}\label{eq:betarot_gauss}
\beta_4^{\mathrm{rot}}=\exp\left(-8\sigma_{\hat{\alpha}^{\mathrm{rot}}}^2\right).
\end{equation}  

The errors on the rotation estimator depend on functions $F_n$ of the intrinsic ellipticity distribution. These are functions of the intrinsic ellipticities and depend upon the distribution of errors on $\bm{\epsilon}^{\mathrm{obs}}-\hat{\bm{\gamma}}$, which provide an estimate of $\bm{\epsilon}^{\mathrm{int}}$. We express estimates of $\bm{\epsilon}^{\mathrm{int}}$ as $\hat{\bm{\epsilon}}^{\mathrm{int}}=\bm{\epsilon}^{\mathrm{obs}}-\hat{\bm{\gamma}}$ and errors on $\hat{\bm{\epsilon}}^{\mathrm{int}}$ as $\bm{\delta}=\hat{\bm{\epsilon}}^{\mathrm{int}}-\bm{\epsilon}^{\mathrm{int}}$.
If the errors on $\hat{\bm{\epsilon}}^{\mathrm{int}}$ are Gaussian distributed with dispersion parameter $\sigma_{\mathrm{\delta}}$, then the $F_n$ functions are given by
\begin{align}\label{eq:F_gauss}
F_n&\left(\left|\bm{\epsilon}^{\mathrm{int}}\right|\right)=\frac{1}{K\left(\left|\bm{\epsilon}^{\mathrm{int}}\right|\right)}\int_0^{\left|\bm{\epsilon}_{\mathrm{max}}^{\mathrm{int}}\right|}\mathrm{d}\left|\hat{\bm{\epsilon}}^{\mathrm{int}}\right|\left|\hat{\bm{\epsilon}}^{\mathrm{int}}\right|\nonumber\\
&\times\exp\left(-\frac{\left|\bm{\epsilon}^{\mathrm{int}}\right|^2+\left|\hat{\bm{\epsilon}}^{\mathrm{int}}\right|^2}{2\sigma_{\delta}^2}\right)I_n\left(\frac{\left|\bm{\epsilon}^{\mathrm{int}}\right|\left|\hat{\bm{\epsilon}}^{\mathrm{int}}\right|}{\sigma_{\delta}^2}\right),
\end{align}
where the integral is carried out over the estimated intrinsic ellipticities and $I_n$ is the $n^{\mathrm{th}}$ order modified Bessel function of the first kind. The error on the components of $\hat{\bm{\epsilon}}^{\mathrm{int}}$ is\footnote{Equation (\ref{eq:err_eps}) ignores correlations between errors on the ellipticities and errors on the shear estimates. For our zero signal simulations and ignoring errors on $\alpha^{\mathrm{int}}$, this approximation is correct to within 5\% for $N=20$ and $\sigma=0.3/\sqrt{2}$. Including errors on $\alpha^{\mathrm{int}}$ and increasing galaxy numbers reduces correlations and improves the accuracy of the approximation.}
\begin{align}\label{eq:err_eps}
\sigma_{\delta}\approx\sqrt{\sigma^2+\sigma_{\hat{\gamma}}^2}.
\end{align}
The constant $K$ is a normalisation constant that depends on $\left|\bm{\epsilon}^{\mathrm{int}}\right|$. It can be shown that $K$ is
\begin{align}\label{eq:K_gauss}
K\left(\left|\epsilon^{\mathrm{int}}\right|\right)&=\int_0^{\left|\bm{\epsilon}_{\mathrm{max}}^{\mathrm{int}}\right|}\mathrm{d}\left|\hat{\bm{\epsilon}}^{\mathrm{int}}\right|\left|\hat{\bm{\epsilon}}^{\mathrm{int}}I_0\left(\frac{\left|\bm{\epsilon}^{\mathrm{int}}\right|\left|\hat{\bm{\epsilon}}^{\mathrm{int}}\right|}{\sigma_{\delta}^2}\right)\right|\nonumber\\
&\times\exp\left(-\frac{\left|\bm{\epsilon}^{\mathrm{int}}\right|^2+\left|\hat{\bm{\epsilon}}^{\mathrm{int}}\right|^2}{2\sigma_{\delta}^2}\right).
\end{align}
The expectation values of the $F_n$ functions are taken over the intrinsic ellipticity distribution, $f_{\mathrm{int}}\left(\left|\bm{\epsilon}^{\mathrm{int}}\right|\right)$
\begin{equation}\label{eq:exp_F}
\left<F_n\right>=\int_0^{\left|\bm{\epsilon}_{\mathrm{max}}^{\mathrm{int}}\right|}\mathrm{d}\left|\bm{\epsilon}^{\mathrm{int}}\right|F_n\left(\left|\bm{\epsilon}^{\mathrm{int}}\right|\right)f_{\mathrm{int}}\left(\left|\bm{\epsilon}^{\mathrm{int}}\right|\right).
\end{equation}

The error on $\hat{\alpha}^{\mathrm{rot}}$ is given in terms of the $F_N$ functions by
\begin{equation}\label{eq:err_alpharot}
\sigma_{\hat{\alpha}^{\mathrm{rot}}}=\frac{\sigma_S}{2\beta_2^{\mathrm{int}}\left<F_1\right>},
\end{equation}
where 
\begin{equation}
\label{eq:disp_trigs}
\sigma^2_{S}=\frac{1}{2N}\left(1-\beta_4^{\mathrm{int}}\left<F_2\right>\right)\text{.}
\end{equation}
Notice that, just as the error in the shear estimator depends on the error in the rotation estimator, the error in the rotation estimator depends on the error in the shear through $\bm{\delta}$ in the $F_n$ functions.

\subsection{Algorithm to approximate the errors}
Similarly to when applying the estimators themselves, the interconnection between the errors can be dealt with via an iterative process. To recover a first estimate of the error on the shear, we can assume that $\beta_4^{\mathrm{rot}}=1$. 
Using equation (\ref{eq:err_eps}), we can calculate the required $F_n$ functions, and hence the error on the rotation estimator. We can then compute an improved estimate of $\sigma_{\mathrm{\hat{\gamma}}}$ using the new value of $\sigma_{\hat{\alpha}^{\mathrm{rot}}}$ in equation (\ref{eq:betarot_gauss}). The above procedure can then be iterated using the improved value of $\sigma_{\mathrm{\hat{\gamma}}}$ as the input when calculating the $F_n$ functions, until successive values of $\sigma_{\mathrm{\hat{\gamma}}}$ are consistent.

\begin{figure}
\includegraphics[width=3.3in]{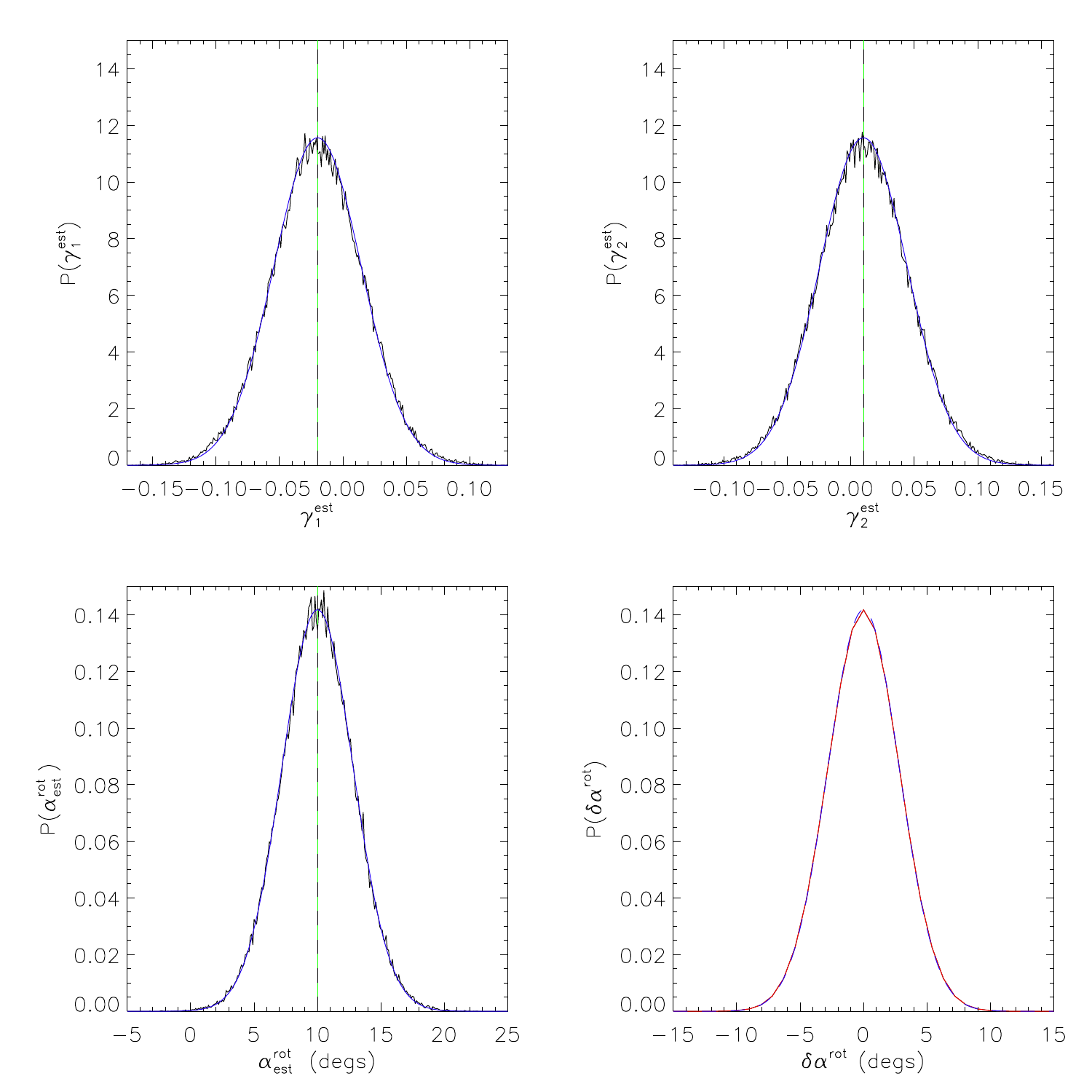}
\caption{Recovered shear and rotation estimates from $10^5$ realizations. Over-plotted in red are the predicted distributions calculated using the procedure described in the text. The bottom right-hand panel shows the error distribution for the rotation using equation (\ref{eq:marg_dt}) (red curve) and the Gaussian approximation (equation (\ref{eq:marg_gauss})) (blue dashed curve).}
\label{fig:err_dt}
\end{figure}

A summary of the steps in the proposed procedure using the approximate Gaussian distribution for $\delta\alpha^{\mathrm{rot}}$ is as follows,
\begin{enumerate}
\item Calculate $\beta_{2}^{\mathrm{int}}$ and $\beta_{4}^{\mathrm{int}}$  using equation (\ref{eq:beta_gauss} (or equation (\ref{eq:beta}) if the errors on $\alpha^{\mathrm{int}}$ are not Gaussian).
\item Calculate the errors on the shear using equation (\ref{eq:err_cbb_dt2}) and assuming that $\beta_4^{\mathrm{rot}}=1$.
\item \label{item:2}Using this error, calculate $\sigma_{\delta}$ with equation (\ref{eq:err_eps}).
\item Use $\sigma_{\delta}$ to construct the $F_{1}$ and $F_{2}$ functions with equation (\ref{eq:F_gauss}) (or equation (\ref{eq:F_n}) if the errors on $\hat{\bm{\epsilon}}^{\mathrm{int}}$ are not Gaussian).
\item Calculate the expectation values $\left<F_{1}\right>$ and $\left<F_{2}\right>$ by integrating over the intrinsic ellipticity distribution (equation (\ref{eq:exp_F})).
\item Estimate the error on $\hat{\alpha}^{\mathrm{rot}}$ using equation (\ref{eq:err_alpharot}).
\item Calculate $\beta_4^{\mathrm{rot}}$ using equation (\ref{eq:betarot_gauss}).
\item \label{item:last}Update the errors on the shear using the new $\beta_4^{\mathrm{rot}}$.
\item Iterate steps \ref{item:2} - \ref{item:last} if required.
\end{enumerate}

Figure \ref{fig:err_dt} shows the shear and rotation estimates recovered from $10^5$ realizations using an input of $\gamma_1=-0.02$, $\gamma_2=0.01$, $\gamma_1^{\mathrm{IA}}=0.001$, $\gamma_2^{\mathrm{IA}}=-0.002$ and $\alpha^{\mathrm{rot}}=10^{\circ}$. We have assumed Gaussian errors of $\sigma_{\alpha^{\mathrm{int}}}=10^{\circ}$ and $\sigma=0.15/\sqrt{2}$. There are 50 galaxies in each realization. We assumed the shear estimates to be approximately Gaussian distributed and, as the errors on the observed ellipticities are also Gaussian, we used equation (\ref{eq:F_gauss}) to calculate the $F_n$ functions. Over-plotted in blue on each histogram are the error distributions predicted using the above procedure. We see that the Gaussian distribution again provides a good fit to the shear and rotation estimates. One iteration was used here, and the difference between the shear errors from the updated zeroth and updated first iteration was less than $10^{-4}$. We also found that the dispersion given in equation (\ref{eq:err_alpharot}) matched the dispersion given by the full distribution of equation (\ref{eq:marg_dt}) to within $0.1^{\circ}$. The bottom right-hand panel shows a comparison of the full distribution for $f_{\delta\alpha^{\mathrm{rot}}}$ and the Gaussian approximation given in equation (\ref{eq:marg_gauss}). The method presented here provided estimates of errors on the shear within 3\% of the values recovered from the simulations shown in Figure \ref{fig:err_dt}. The estimate of the error on the rotation was within 1\%.

\subsection{Simplification to the algorithm}
If we assume that for a specific source errors on $\hat{\bm{\epsilon}}^{\mathrm{int}}$ are much less than $|\bm{\epsilon}^{\mathrm{int}}|$, that is, if $\sigma_{\delta}\ll|\bm{\epsilon}^{\mathrm{int}}|$, then it can be shown that $F_n\left(\left|\bm{\epsilon}^{\mathrm{int}}\right|\right)\approx1$. If we further assume that the errors are much less than a typical source ellipticity, so that $\sigma_{\mathrm{\delta}}\ll\sigma_{\epsilon}$,  we can write $\left<F_n\left(\left|\bm{\epsilon}^{\mathrm{int}}\right|\right)\right>\approx1$. This allows us to write the error on $\hat{\alpha}_{\mathrm{rot}}$ as
\begin{equation}\label{eq:simp_sigrot}
\sigma_{\hat{\alpha}^{\mathrm{rot}}}=\frac{1}{2\beta_2^{\mathrm{int}}}\sqrt{\frac{1-\beta^\text{int}_4}{2N}}.
\end{equation}
The error on the shear can then be calculated using equation (\ref{eq:err_cbb_dt2}). If the errors on the intrinsic orientation are Gaussian distributed, we can simplify the $\beta_n^{\mathrm{int}}$ terms to give
\begin{equation}\label{eq:simplest_sigrot}
\sigma_{\hat{\alpha}^{\mathrm{rot}}}=\sqrt{\frac{\sinh\left(4\sigma_{\alpha_{\mathrm{rms}}}^2\right)}{4N}}.
\end{equation}

\begin{figure}
\includegraphics[width=3.3in]{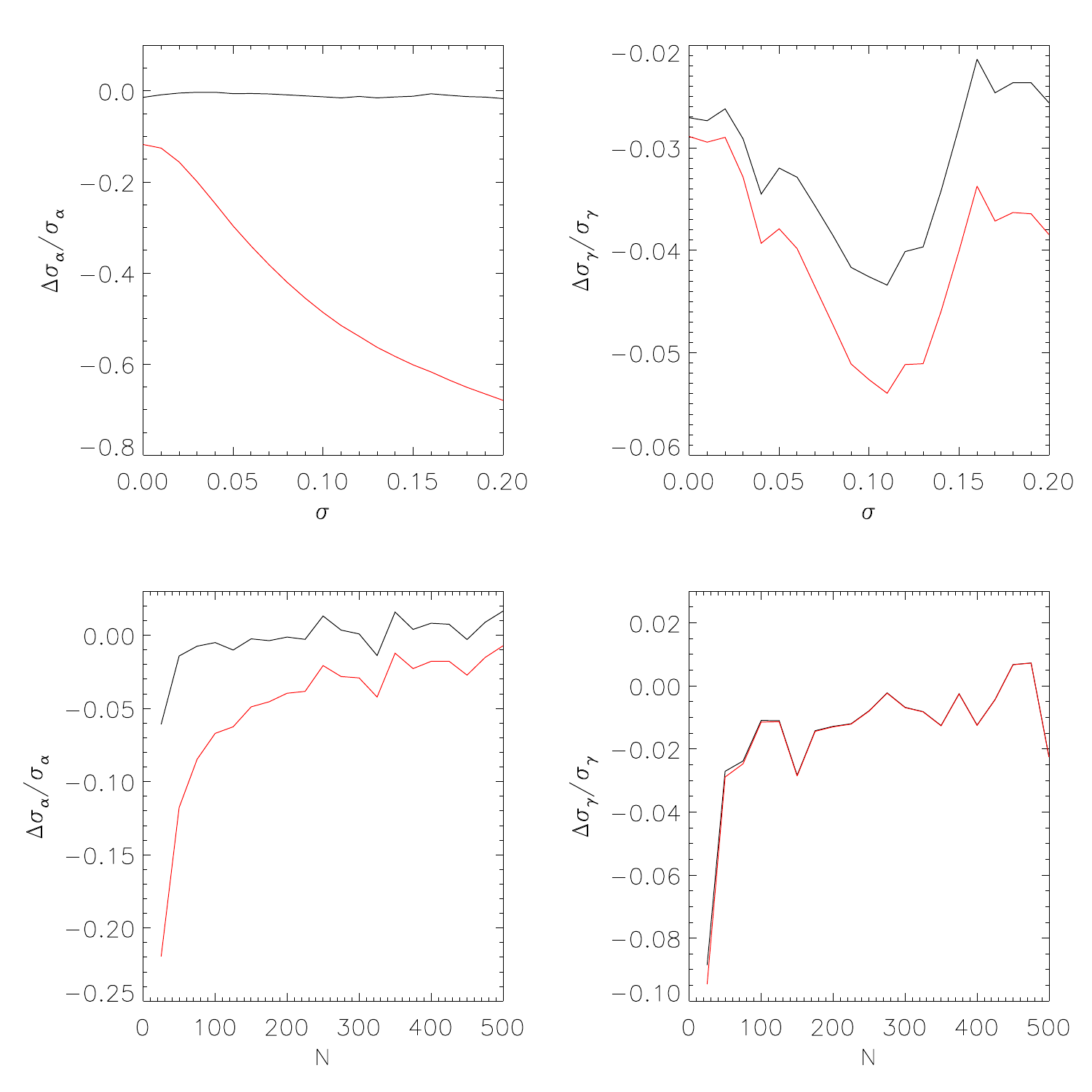}
\caption{A comparison of the performances of the full iterative approach to estimating the errors (black curves) and the simplified approach using equation (\ref{eq:simplest_sigrot}) (red curves). In each case, we show the fractional differences between the estimated errors and those recovered from $10^4$ Monte-Carlo simulations. The top panels show the errors as a function of the errors on the ellipticity measurements, with the number of galaxies fixed to $N=50$. The bottom panels show the errors as a function of galaxy number with the ellipticity measurement error fixed to zero.}
\label{fig:simp_sig}
\end{figure}

Figure \ref{fig:simp_sig} shows the fractional differences between errors on the rotation and shear estimates calculated using the full iterative approach discussed above and the errors recovered from $10^4$ Monte-Carlo simulations (black curves), where we have assumed zero input shear and rotation signals, and a Gaussian error on the rotation of $10^{\circ}$. In addition, the red curves show the fractional differences between the errors calculated using the simplified form of equation (\ref{eq:simplest_sigrot}) to estimate $\sigma_{\hat{\alpha}^{\mathrm{rot}}}$ and those from the simulations.

We see that for the full iterative approach with $N=50$, the recovered errors on the shear and rotation agree with the simulations to within a few percent over the full range of considered $\sigma$ values. The simplified method deviates from the simulations for large $\sigma$ and agrees at $\sim$$50\%$ when $\sigma\sim0.1$. When using the simplified method to estimate the error on the shear (the red curve in top-right panel of Figure \ref{fig:simp_sig}), the errors on the shear agree with those from the simulations to within a few percent. This is indicative of the fact that, in this case, errors on the shear estimates are dominated by measurement errors on $\alpha^{\mathrm{int}}$ and shape noise.

As a function of galaxy number with $\sigma=0$, the full iterative approach for estimating the rotation error performs well over the full range, whereas the simplified approach shows a small deviation at low numbers. Both the full and simplified approaches display a similar performance when estimating the shear errors, deviating by $\sim$10\% when $N=25$.

\subsection{Derivation of equations}
\label{app_deriv}
Errors on the rotation estimator in equation (\ref{eq:rot_est}) arise from the propagation of errors on the means of the trigonometric functions. Let us assume that there are enough galaxies in the sample that errors on the means are Gaussian distributed as a consequence of the central limit theorem. Assuming zero shear and IA, the error distribution on $\hat{\alpha}^{\mathrm{rot}}$ must be independent of $\alpha^{\mathrm{int}}$. Therefore, if we set $\alpha^{\mathrm{int}}$ to zero, the expectation values of the means are 
\begin{align}\label{eq:mean_trigs}
\left<C\right>=&\beta_2^{\mathrm{int}}\left<F_1\left(\left|\bm{\epsilon}^{\mathrm{int}}\right|\right)\right>,\nonumber\\
\left<S\right>=&0,
\end{align}
where $C=\frac{1}{N}\sum_{i=1}^{N}\cos\left(2\alpha_{i}^{\mathrm{new}}-2\hat{\alpha}_{i}^{\mathrm{int}}\right)$ and $S=\frac{1}{N}\sum_{i=1}^{N}\sin\left(2\alpha_{i}^{\mathrm{new}}-2\hat{\alpha}_{i}^{\mathrm{int}}\right)$. Note that when the errors on the intrinsic orientations ($\delta\alpha^\text{int}$) recovered from the polarisation data are not Gaussian (as was assumed in equation (\ref{eq:beta_gauss})), the $\beta^\text{int}_n$ functions are given more generally by
\begin{equation}\label{eq:beta} 
\beta_n^{\mathrm{int}}=\left<\cos\left(n\delta\alpha^{\mathrm{int}}\right)\right>.
\end{equation}
The terms $F_n$ are defined in \cite{whittaker2014}, but in this case, they are functions of the intrinsic ellipticities and depend upon the distribution of errors on $\bm{\epsilon}^{\mathrm{obs}}-\hat{\bm{\gamma}}$, which provide an estimate of $\bm{\epsilon}^{\mathrm{int}}$ and have a contribution from errors on the measurements of $\bm{\epsilon}^{\mathrm{obs}}$ and the estimate of $\bm{\gamma}$. Let us express estimates of $\bm{\epsilon}^{\mathrm{int}}$ as
\begin{equation}\label{eq:est_int}
\hat{\bm{\epsilon}}^{\mathrm{int}}=\bm{\epsilon}^{\mathrm{obs}}-\hat{\bm{\gamma}},
\end{equation}
and errors on $\hat{\bm{\epsilon}}^{\mathrm{int}}$ in polar form as
\begin{equation}\label{eq:err_polar}
\bm{\delta}=\left|\bm{\delta}\right|\exp\left(2i\alpha^{\mathrm{err}}\right)\equiv\hat{\bm{\epsilon}}^{\mathrm{int}}-\bm{\epsilon}^{\mathrm{int}}.
\end{equation}
The $F_n$ functions, in this case, are then
\begin{align}\label{eq:F_n}
F_n\left(\left|\bm{\epsilon}^{\mathrm{int}}\right|\right)=&\frac{1}{\pi}\int_{-\frac{\pi}{2}}^{\frac{\pi}{2}}\int_0^{\left|\bm{\delta}_{\mathrm{max}}\right|\left(\alpha^{\mathrm{err}}\right)}\mathrm{d}\left|\bm{\delta}\right|\mathrm{d}\alpha^{\mathrm{err}}f_{\mathrm{err}}\left(\left|\bm{\delta}\right|,\alpha^{\mathrm{err}}\right)\nonumber\\
&\times g_n\left(\left|\bm{\epsilon}^{\mathrm{int}}\right|,\left|\bm{\delta}\right|,\alpha^{\mathrm{err}}\right),
\end{align}
where $f_{\mathrm{err}}$ is the distribution of errors on $\hat{\bm{\epsilon}}^{\mathrm{int}}$ and $\left|\bm{\delta}_{\mathrm{max}}\right|\left(\alpha^{\mathrm{err}}\right)$ is the maximum error, which has been expressed as a function of the angle $\alpha^{\mathrm{err}}$ due to a possible asymmetry in the error distribution imposed by the maximum intrinsic ellipticity, $\left|\bm{\epsilon}^{\mathrm{int}}_{\mathrm{max}}\right|$. The function $g_n\left(\left|\bm{\epsilon}^{\mathrm{int}}\right|,\left|\bm{\delta}\right|,\alpha^{\mathrm{err}}\right)$ is given as
\begin{equation}\label{eq:g1_F}
g_n\left(\left|\bm{\epsilon}^{\mathrm{int}}\right|,\left|\bm{\delta}\right|,\alpha^{\mathrm{err}}\right)=\cos\left(2n\alpha'\right),
\end{equation}
with
\begin{align}\label{eq:edash}
\left|\bm{\epsilon'}\right|\cos\left(2\alpha'\right)=&\left|\bm{\epsilon}^{\mathrm{int}}\right|+\left|\bm{\delta}\right|\cos\left(2\alpha^{\mathrm{err}}\right), \nonumber \\
\left|\bm{\epsilon'}\right|\sin\left(2\alpha'\right)=&\left|\bm{\delta}\right|\sin\left(2\alpha^{\mathrm{err}}\right). \nonumber \\
\end{align}

If the errors on $\hat{\bm{\epsilon}}^{\mathrm{int}}$ are Gaussian distributed with dispersion parameter $\sigma_{\mathrm{\delta}}$, it can be shown that
\begin{align}\label{eq:F_gauss2}
F_n\left(\left|\bm{\epsilon}^{\mathrm{int}}\right|\right)=&\frac{1}{K\left(\left|\bm{\epsilon}^{\mathrm{int}}\right|\right)}\int_0^{\left|\hat{\bm{\epsilon}}_{\mathrm{max}}^{\mathrm{int}}\right|}\mathrm{d}\left|\hat{\bm{\epsilon}}^{\mathrm{int}}\right|\left|\hat{\bm{\epsilon}}^{\mathrm{int}}\right|\nonumber\\
&\times\exp\left(-\frac{\left|\bm{\epsilon}^{\mathrm{int}}\right|^2+\left|\hat{\bm{\epsilon}}^{\mathrm{int}}\right|^2}{2\sigma_{\delta}^2}\right)I_n\left(\frac{\left|\bm{\epsilon}^{\mathrm{int}}\right|\left|\hat{\bm{\epsilon}}^{\mathrm{int}}\right|}{\sigma_{\delta}^2}\right),
\end{align}
where the integral is now carried out over the estimated intrinsic ellipticities and $I_n$ is the $n^{\mathrm{th}}$ order modified Bessel function of the first kind. $K$ is a normalization constant which depends on $\left|\bm{\epsilon}^{\mathrm{int}}\right|$. This dependency is due to the fact that we cannot have any estimated intrinsic ellipticity greater than $\left|\hat{\bm{\epsilon}}_{\mathrm{max}}^{\mathrm{int}}\right|$, and we also require that the error distribution be circularly symmetric so that $\hat{\bm{\epsilon}}^{\mathrm{int}}$ is unbiased. It can be shown that $K$ is
\begin{align}\label{eq:K_gauss2}
K\left(\left|\epsilon^{\mathrm{int}}\right|\right)=&\int_0^{\left|\hat{\bm{\epsilon}}_{\mathrm{max}}^{\mathrm{int}}\right|}\mathrm{d}\left|\hat{\bm{\epsilon}}^{\mathrm{int}}\right|\left|\hat{\bm{\epsilon}}^{\mathrm{int}}\right|\nonumber\\
&\times\exp\left(-\frac{\left|\bm{\epsilon}^{\mathrm{int}}\right|^2+\left|\hat{\bm{\epsilon}}^{\mathrm{int}}\right|^2}{2\sigma_{\delta}^2}\right)I_0\left(\frac{\left|\bm{\epsilon}^{\mathrm{int}}\right|\left|\hat{\bm{\epsilon}}^{\mathrm{int}}\right|}{\sigma_{\delta}^2}\right).
\end{align}

The variances and covariance of the mean trigonometric functions are given by \citep{whittaker2014}
\begin{align}\label{eq:disp_trigs2}
\sigma^2_{C}=&\frac{1}{2N}\left(1+\beta_4^{\mathrm{int}}\left<F_2\right>-2\beta_2^{\mathrm{int}^2}\left<F_1\right>^2\right),\nonumber\\
\sigma^2_{S}=&\frac{1}{2N}\left(1-\beta_4^{\mathrm{int}}\left<F_2\right>\right),\nonumber\\
\mathrm{Cov}\left(C,S\right)=&0,
\end{align}
where $\sigma^2_{C},\sigma^2_{S}\le1/(2N)$. The expectation values of the $F_n$ functions are taken over the intrinsic ellipticity distribution, $f_{\mathrm{int}}\left(\left|\bm{\epsilon}^{\mathrm{int}}\right|\right)$:
\begin{equation}\label{eq:exp_F2}
\left<F_n\right>=\int_0^{\left|\bm{\epsilon}_{\mathrm{max}}^{\mathrm{int}}\right|}\mathrm{d}\left|\bm{\epsilon}^{\mathrm{int}}\right|F_n\left(\left|\bm{\epsilon}^{\mathrm{int}}\right|\right)f_{\mathrm{int}}\left(\left|\bm{\epsilon}^{\mathrm{int}}\right|\right).
\end{equation}
This integral is required since the dispersion of the error on a particular estimate of $\alpha^{\mathrm{new}}$ depends on the ellipticity of the object given the error distribution $f_{\mathrm{err}}$. One must therefore have a prior estimate of the intrinsic ellipticity distribution if one requires an accurate estimate of the error distribution on $\alpha^{\mathrm{rot}}$. 

Now that we know the variances and expectation values of the mean trigonometric functions, we can write the distribution of the estimates in a frame of reference rotated by $45^{\circ}$ using the central limit theorem as
\begin{equation}\label{eq:gauss_clt}
f_{CS}\left(\bm{n}\right)=\frac{1}{2\pi\sigma_C\sigma_S}\exp\left\{-\frac{1}{2}\left(\bm{n}-\bm{n}_0\right)^{\mathrm{T}}\bm{\Sigma}^{-1}\left(\bm{n}-\bm{n}_0\right)\right\},
\end{equation}
with
\begin{equation}\label{eq:deltacs}
\bm{n}=\frac{1}{\sqrt{2}}\left(
\begin{array}{c}
C-S \\
C+S
\end{array} \right),\,\,\,
\bm{n}_0=\frac{1}{\sqrt{2}}\left(
\begin{array}{c}
\beta_2^{\mathrm{int}}\left<F_1\right>\\
\beta_2^{\mathrm{int}}\left<F_1\right>
\end{array} \right),
\end{equation}
and 
\begin{equation}\label{eq:inv_covmat}
\bm{\Sigma}^{-1}=\frac{1}{2}\left(
\begin{array}{cc}
\frac{1}{\sigma_C^2}+\frac{1}{\sigma_S^2} & \frac{1}{\sigma_C^2}-\frac{1}{\sigma_S^2} \\
\frac{1}{\sigma_C^2}-\frac{1}{\sigma_S^2} & \frac{1}{\sigma_C^2}+\frac{1}{\sigma_S^2}
\end{array} \right).
\end{equation}
The rotation has been performed so that we can ignore the effects of $C$ wrapping around, since $-1\le C\le1$. In addition to this rotation, to avoid wrapping, we also require that $N\gtrsim20$. Using simple simulations, we have verified that the error distributions for $C$ and $S$ are well approximated as Gaussian distributions when $N=20$, and the accuracy of this approximation increases as $N$ increases.

Since we are assuming that the rotation is zero, we can write $\bm{n}$ in polar form as
\begin{equation}\label{eq:ndef_Palpha}
\bm{n}=P\left(
\begin{array}{c}
\cos\left(2\delta\alpha^{\mathrm{rot}}+\frac{\pi}{4}\right) \\
\sin\left(2\delta\alpha^{\mathrm{rot}}+\frac{\pi}{4}\right)
\end{array} \right),
\end{equation}
and the distribution of $\delta\alpha^{\mathrm{rot}}$ is therefore
\begin{align}
f_{\delta\alpha^{\mathrm{rot}}}&\left(\delta\alpha^{\mathrm{rot}}\right)=2\int_{0}^{\infty}\mathrm{d}PPf_{CS}\left(P,\delta\alpha^{\mathrm{rot}}\right).
\end{align}
Carrying out this marginalization gives 
\begin{equation}\label{eq:marg_dt}
f_{\delta\alpha^{\mathrm{rot}}}\left(\delta\alpha^{\mathrm{rot}}\right)=\frac{K}{2}\left[\frac{\exp\left(-AG^2\right)}{A}+G\sqrt{\frac{\pi}{A}}\left(1+\mathrm{erf}\left(\sqrt{A}G\right)\right)\right],
\end{equation}
where
\begin{align}\label{eq:marg_agk}
A=&\frac{\cos^2\left(2\delta\alpha^{\mathrm{rot}}\right)\sigma_S^2+\sin^2\left(2\delta\alpha^{\mathrm{rot}}\right)\sigma_C^2}{2\sigma_C^2\sigma_S^2},\nonumber\\
G=&\frac{\beta_2^{\mathrm{int}}\left<F_1\right>\cos\left(2\delta\alpha^{\mathrm{rot}}\right)\sigma_S^2}{\cos^2\left(2\delta\alpha^{\mathrm{rot}}\right)\sigma_S^2+\sin^2\left(2\delta\alpha^{\mathrm{rot}}\right)\sigma_C^2},\nonumber\\
K=&\frac{1}{\pi\sigma_C\sigma_S}\exp\biggl\{\nonumber\\
&-\frac{1}{2\sigma_c^2\sigma_S^2}\left[\beta_2^{\mathrm{int}^2}\left<F_1\right>^2\sigma_S^2-G\beta_2^{\mathrm{int}}\left<F_1\right>\cos\left(2\delta\alpha^{\mathrm{rot}}\right)\sigma_S^2\right]\biggr\}.
\end{align}
Assuming that $\sigma_C,\sigma_S\ll\beta_2^{\mathrm{int}}\left<F_1\right>$ so that $\delta\alpha^{\mathrm{rot}}\ll1$, we can simplify equation (\ref{eq:marg_agk}) to leading order in $\delta\alpha^{\mathrm{rot}}$ to give
\begin{align}
A\approx &\frac{1}{2\sigma_C^2},\nonumber\\
G\approx &\beta_2^{\mathrm{int}}\left<F_1\right>,\nonumber\\
K\approx &\frac{1}{\pi\sigma_C\sigma_S}\exp\left(-\frac{2\beta_2^{\mathrm{int}^2}\left<F_1\right>^2\delta\alpha^{\mathrm{rot}^2}}{\sigma_S^2}\right).
\end{align}
Substituting these simplifications into equation (\ref{eq:marg_dt}), we find that the distribution of $\delta\alpha^{\mathrm{rot}}$ is approximately Gaussian, with 
\begin{equation}\label{eq:marg_gauss}
f_{\delta\alpha^{\mathrm{rot}}}\left(\delta\alpha^{\mathrm{rot}}\right)=\sqrt{\frac{2\beta_2^{\mathrm{int}^2}\left<F_1\right>^2}{\pi\sigma_S^2}}\exp\left(-\frac{2\beta_2^{\mathrm{int}^2}\left<F_1\right>^2\delta\alpha^{\mathrm{rot}^2}}{\sigma_S^2}\right).
\end{equation}
The error on $\hat{\alpha}^{\mathrm{rot}}$ can now be read directly from equation (\ref{eq:marg_gauss}) and is
\begin{equation}\label{eq:err_alpharot2}
\sigma_{\hat{\alpha}^{\mathrm{rot}}}=\frac{\sigma_S}{2\beta_2^{\mathrm{int}}\left<F_1\right>}.
\end{equation}

In order to construct the distribution in equation (\ref{eq:marg_dt}), one must calculate $\left<F_1\right>$, $\sigma_C$ and $\sigma_S$. These require a knowledge of the errors on the terms $\hat{\bm{\epsilon}}^{\mathrm{int}}$. Errors on these terms have a contribution from errors on the ellipticities and the shear. Following the approach of \cite{whittaker2015}, the error on the shear can be approximated as 
\begin{equation}\label{eq:err_cbb_dt}
\sigma_{\hat{\gamma}}^2=\frac{1}{N\beta_4^{\mathrm{int}^2}}\left[\left(1+\beta_4^{\mathrm{int}^2}-2\beta_4^{\mathrm{int}}\beta_4^{\mathrm{tot}}\right)\sigma_{\epsilon}^2+\left(1+\beta_4^{\mathrm{int}^2}\right)\sigma^2\right],
\end{equation}
where $\sigma$ is the error on the ellipticities and, assuming a large enough sample size,
\begin{align}\label{eq:err_betat}
\beta_4^{\mathrm{tot}}=&\left<\left<\cos\left(4\delta\alpha^{\mathrm{int}}+4\delta\alpha^{\mathrm{rot}}\right)\right>\right>,\nonumber\\
=&\left<\left<\cos\left(4\delta\alpha^{\mathrm{int}}\right)\right>\cos\left(4\delta\alpha^{\mathrm{rot}}\right)\right>,\nonumber\\
\approx&\beta_4^{\mathrm{int}}\beta_4^{\mathrm{rot}},
\end{align}
where
\begin{equation}\label{eq:betarot}
\beta_4^{\mathrm{rot}}=\left<\cos\left(4\delta\alpha^{\mathrm{rot}}\right)\right>. 
\end{equation}
Therefore, we can write
\begin{equation}\label{eq:err_cbb_dt2_2}
\sigma_{\hat{\gamma}}^2=\frac{1}{N\beta_4^{\mathrm{int}^2}}\left\{\left[1+\beta_4^{\mathrm{int}^2}\left(1-2\beta_4^{\mathrm{rot}}\right)\right]\sigma_{\epsilon}^2+\left(1+\beta_4^{\mathrm{int}^2}\right)\sigma^2\right\}.
\end{equation}
If the error distribution $f_{\delta\alpha^{\mathrm{rot}}}$ is Gaussian, then, using equation (\ref{eq:beta_gauss}), $\beta_4^{\mathrm{rot}}$ is 
\begin{equation}\label{eq:betarot_gauss2}
\beta_4^{\mathrm{rot}}=\exp\left(-8\sigma_{\hat{\alpha}^{\mathrm{rot}}}^2\right).
\end{equation}
A finite number of galaxies in the sample implies that there will be correlations between $\left<\cos\left(4\delta\alpha^{\mathrm{int}}\right)\right>$ and $\cos\left(4\delta\alpha^{\mathrm{rot}}\right)$ on the second line of equation (\ref{eq:err_betat}). Using simulations similar to those used in section \ref{sec_properties} but with zero signal, including Gaussian errors on $\alpha^{\mathrm{int}}$ with standard deviation $\alpha_\text{rms}=10^{\circ}$ and ignoring errors on the observed ellipticities, we find that, for $N=20$ galaxies, the approximation in equation (\ref{eq:err_betat}) is correct to $<$0.1\%. Including errors on the ellipticity measurements and increasing the number of galaxies further reduces correlations and improves the accuracy of the approximation.

\bibliographystyle{mn2e}
\bibliography{rotest}

\begin{thebibliography}{}

\bibitem[\protect\citeauthoryear{{Adamek}, {Durrer} \& {Kunz}}{{Adamek}
  et~al.}{2014}]{adamek2014}
{Adamek} J.,  {Durrer} R.,    {Kunz} M.,  2014, Classical and Quantum Gravity,
  31, 234006

\bibitem[\protect\citeauthoryear{Amendola et~al.,}{Amendola
  et~al.}{2016}]{euclid}
Amendola L.,  et~al., 2016

\bibitem[\protect\citeauthoryear{{Amendola}}{{Amendola}}{2013}]{euclid3}
{Amendola} L. e.~a.,  2013, Living Reviews in Relativity, 16

\bibitem[\protect\citeauthoryear{{Bacon}, {Goldberg}, {Rowe} \&
  {Taylor}}{{Bacon} et~al.}{2006}]{bacon2006}
{Bacon} D.~J.,  {Goldberg} D.~M.,  {Rowe} B.~T.~P.,    {Taylor} A.~N.,  2006,
  \mnras, 365, 414

\bibitem[\protect\citeauthoryear{{Bacon}, {Refregier} \& {Ellis}}{{Bacon}
  et~al.}{2000}]{bacon2000}
{Bacon} D.~J.,  {Refregier} A.~R.,    {Ellis} R.~S.,  2000, \mnras, 318, 625

\bibitem[\protect\citeauthoryear{Battye, Garbrecht \& Moss}{Battye
  et~al.}{2006}]{battye2006}
Battye R.~A.,  Garbrecht B.,    Moss A.,  2006, JCAP, 0609, 007

\bibitem[\protect\citeauthoryear{{Becker}, {White} \& {Helfand}}{{Becker}
  et~al.}{1995}]{becker1995}
{Becker} R.~H.,  {White} R.~L.,    {Helfand} D.~J.,  1995, \apj, 450, 559

\bibitem[\protect\citeauthoryear{{Bernardeau}, {Bonvin} \&
  {Vernizzi}}{{Bernardeau} et~al.}{2010}]{bernardeau2010}
{Bernardeau} F.,  {Bonvin} C.,    {Vernizzi} F.,  2010, \prd, 81, 083002

\bibitem[\protect\citeauthoryear{{Booth}, {de Blok}, {Jonas} \&
  {Fanaroff}}{{Booth} et~al.}{2009}]{booth2009}
{Booth} R.~S.,  {de Blok} W.~J.~G.,  {Jonas} J.~L.,    {Fanaroff} B.,  2009,
  ArXiv e-prints

\bibitem[\protect\citeauthoryear{{Bridle} \& {King}}{{Bridle} \&
  {King}}{2007}]{bridle2007}
{Bridle} S.,  {King} L.,  2007, New Journal of Physics, 9, 444

\bibitem[\protect\citeauthoryear{{Brown}, {Bacon}, {Camera}, {Harrison},
  {Joachimi}, {Metcalf}, {Pourtsidou}, {Takahashi}, {Zuntz}, {Abdalla},
  {Bridle}, {Jarvis}, {Kitching}, {Miller} \& {Patel}}{{Brown}
  et~al.}{2015}]{brown2015}
{Brown} M.~L.,  {Bacon} D.~J.,  {Camera} S.,  {Harrison} I.,  {Joachimi} B.,
  {Metcalf} R.~B.,  {Pourtsidou} A.,  {Takahashi} K.,  {Zuntz} J.~A.,
  {Abdalla} F.~B.,  {Bridle} S.,  {Jarvis} M.,  {Kitching} T.~D.,  {Miller} L.,
     {Patel} P.,  2015, preprint (arXiv:1501.03828)

\bibitem[\protect\citeauthoryear{{Brown} \& {Battye}}{{Brown} \&
  {Battye}}{2011}]{brown2011}
{Brown} M.~L.,  {Battye} R.~A.,  2011, \mnras, 410, 2057

\bibitem[\protect\citeauthoryear{Broyden}{Broyden}{1965}]{broyden1965}
Broyden C.,  1965, Mathematics of Computation, 19, 577

\bibitem[\protect\citeauthoryear{{Bruni}, {Thomas} \& {Wands}}{{Bruni}
  et~al.}{2014}]{bruni2014}
{Bruni} M.,  {Thomas} D.~B.,    {Wands} D.,  2014, \prd, 89, 044010

\bibitem[\protect\citeauthoryear{{Camera}, {Harrison}, {Bonaldi} \&
  {Brown}}{{Camera} et~al.}{2016}]{camera2016}
{Camera} S.,  {Harrison} I.,  {Bonaldi} A.,    {Brown} M.~L.,  2016, ArXiv
  e-prints

\bibitem[\protect\citeauthoryear{{Chang}, {Refregier} \& {Helfand}}{{Chang}
  et~al.}{2004}]{chang2004}
{Chang} T.,  {Refregier} A.,    {Helfand} D.~J.,  2004, \apj, 617, 794

\bibitem[\protect\citeauthoryear{Contaldi, Hindmarsh \& Magueijo}{Contaldi
  et~al.}{1999}]{contaldi1998}
Contaldi C.,  Hindmarsh M.,    Magueijo J.,  1999, Phys. Rev. Lett., 82, 2034

\bibitem[\protect\citeauthoryear{Copeland, Myers \& Polchinski}{Copeland
  et~al.}{2004}]{copeland2003}
Copeland E.~J.,  Myers R.~C.,    Polchinski J.,  2004, JHEP, 06, 013

\bibitem[\protect\citeauthoryear{{Crittenden}, {Natarajan}, {Pen} \&
  {Theuns}}{{Crittenden} et~al.}{2001}]{crittenden2001}
{Crittenden} R.~G.,  {Natarajan} P.,  {Pen} U.~L.,    {Theuns} T.,  2001, \apj,
  559, 552

\bibitem[\protect\citeauthoryear{Demetroullas \& Brown}{Demetroullas \&
  Brown}{2016}]{demetroullas2015}
Demetroullas C.,  Brown M.~L.,  2016, Mon. Not. Roy. Astron. Soc., 456, 3100

\bibitem[\protect\citeauthoryear{{di Serego Alighieri}}{{di Serego
  Alighieri}}{2011}]{1011.4865}
{di Serego Alighieri} S.,  2011, Astrophysics and Space Science Proceedings,
  22, 139

\bibitem[\protect\citeauthoryear{{Dodelson}}{{Dodelson}}{2003}]{dodelsonbook}
{Dodelson} S.,  2003, {Modern cosmology}

\bibitem[\protect\citeauthoryear{{Dodelson}, {Kolb}, {Matarrese}, {Riotto} \&
  {Zhang}}{{Dodelson} et~al.}{2005}]{dodelson2005}
{Dodelson} S.,  {Kolb} E.~W.,  {Matarrese} S.,  {Riotto} A.,    {Zhang} P.,
  2005, \prd, 72, 103004

\bibitem[\protect\citeauthoryear{{Dodelson}, {Rozo} \& {Stebbins}}{{Dodelson}
  et~al.}{2003}]{dodelson2003}
{Dodelson} S.,  {Rozo} E.,    {Stebbins} A.,  2003, Physical Review Letters,
  91, 021301

\bibitem[\protect\citeauthoryear{Dvali \& Vilenkin}{Dvali \&
  Vilenkin}{2004}]{dvali2004}
Dvali G.,  Vilenkin A.,  2004, JCAP, 0403, 010

\bibitem[\protect\citeauthoryear{{Dyer} \& {Shaver}}{{Dyer} \&
  {Shaver}}{1992}]{dyer1992}
{Dyer} C.~C.,  {Shaver} E.~G.,  1992, \apjl, 390, L5

\bibitem[\protect\citeauthoryear{{Eisenstein} \& {Hu}}{{Eisenstein} \&
  {Hu}}{1999}]{eisenstein1999}
{Eisenstein} D.~J.,  {Hu} W.,  1999, \apj, 511, 5

\bibitem[\protect\citeauthoryear{{Faraoni}}{{Faraoni}}{2008}]{faraoni2008}
{Faraoni} V.,  2008, \na, 13, 178

\bibitem[\protect\citeauthoryear{Fu et~al.,}{Fu  et~al.}{2014}]{cfhtlens}
Fu L.,  et~al., 2014, Mon. Not. Roy. Astron. Soc., 441, 2725

\bibitem[\protect\citeauthoryear{{Galaverni}, {Gubitosi}, {Paci} \&
  {Finelli}}{{Galaverni} et~al.}{2015}]{1411.6287}
{Galaverni} M.,  {Gubitosi} G.,  {Paci} F.,    {Finelli} F.,  2015, \jcap, 8,
  031

\bibitem[\protect\citeauthoryear{{G{\'o}rski}, {Hivon}, {Banday}, {Wandelt},
  {Hansen}, {Reinecke} \& {Bartelmann}}{{G{\'o}rski} et~al.}{2005}]{gorski2005}
{G{\'o}rski} K.~M.,  {Hivon} E.,  {Banday} A.~J.,  {Wandelt} B.~D.,  {Hansen}
  F.~K.,  {Reinecke} M.,    {Bartelmann} M.,  2005, \apj, 622, 759

\bibitem[\protect\citeauthoryear{Harrison, Camera, Zuntz \& Brown}{Harrison
  et~al.}{2016}]{harrison2016}
Harrison I.,  Camera S.,  Zuntz J.,    Brown M.~L.,  2016

\bibitem[\protect\citeauthoryear{Hildebrandt et~al.,}{Hildebrandt
  et~al.}{2016}]{kids450}
Hildebrandt H.,  et~al., 2016

\bibitem[\protect\citeauthoryear{{Hirata} \& {Seljak}}{{Hirata} \&
  {Seljak}}{2003}]{hirata2003}
{Hirata} C.~M.,  {Seljak} U.,  2003, \prd, 68, 083002

\bibitem[\protect\citeauthoryear{{Jain}, {Seljak} \& {White}}{{Jain}
  et~al.}{2000}]{jain2000}
{Jain} B.,  {Seljak} U.,    {White} S.,  2000, \apj, 530, 547

\bibitem[\protect\citeauthoryear{Johnston \& Wall}{Johnston \&
  Wall}{2008}]{johnston2008}
Johnston S.,  Wall J.,  2008, Exper. Astron., 22, 151

\bibitem[\protect\citeauthoryear{{Kaiser}, {Wilson} \& {Luppino}}{{Kaiser}
  et~al.}{2000}]{kaiser2000}
{Kaiser} N.,  {Wilson} G.,    {Luppino} G.~A.,  2000, preprint
  (astro-ph/0003338)

\bibitem[\protect\citeauthoryear{Kibble}{Kibble}{1976}]{kibble1976}
Kibble T. W.~B.,  1976, J. Phys., A9, 1387

\bibitem[\protect\citeauthoryear{{Kirk}, {Brown}, {Hoekstra}, {Joachimi},
  {Kitching}, {Mandelbaum}, {Sif{\'o}n}, {Cacciato}, {Choi}, {Kiessling},
  {Leonard}, {Rassat} \& {Sch{\"a}fer}}{{Kirk} et~al.}{2015}]{kirk2015}
{Kirk} D.,  {Brown} M.~L.,  {Hoekstra} H.,  {Joachimi} B.,  {Kitching} T.~D.,
  {Mandelbaum} R.,  {Sif{\'o}n} C.,  {Cacciato} M.,  {Choi} A.,  {Kiessling}
  A.,  {Leonard} A.,  {Rassat} A.,    {Sch{\"a}fer} B.~M.,  2015, \ssr, 193,
  139

\bibitem[\protect\citeauthoryear{{Krause} \& {Hirata}}{{Krause} \&
  {Hirata}}{2010}]{krause2010}
{Krause} E.,  {Hirata} C.~M.,  2010, \aap, 523, A28

\bibitem[\protect\citeauthoryear{{Kronberg}, {Dyer}, {Burbidge} \&
  {Junkkarinen}}{{Kronberg} et~al.}{1991}]{kronberg1991}
{Kronberg} P.~P.,  {Dyer} C.~C.,  {Burbidge} E.~M.,    {Junkkarinen} V.~T.,
  1991, \apjl, 367, L1

\bibitem[\protect\citeauthoryear{{Laureijs}, {Amiaux}, {Arduini},
  {Augu{\`e}res}, {Brinchmann}, {Cole}, {Cropper}, {Dabin}, {Duvet}, {Ealet} \&
  et al.}{{Laureijs} et~al.}{2011}]{euclid2}
{Laureijs} R.,  {Amiaux} J.,  {Arduini} S.,  {Augu{\`e}res} J.~.,  {Brinchmann}
  J.,  {Cole} R.,  {Cropper} M.,  {Dabin} C.,  {Duvet} L.,  {Ealet} A.,    et
  al. 2011, ArXiv e-prints

\bibitem[\protect\citeauthoryear{{LSST Dark Energy Science
  Collaboration}}{{LSST Dark Energy Science Collaboration}}{2012}]{lsst}
{LSST Dark Energy Science Collaboration} 2012, ArXiv e-prints

\bibitem[\protect\citeauthoryear{Martins \& Shellard}{Martins \&
  Shellard}{1996}]{martins1996}
Martins C. J. A.~P.,  Shellard E. P.~S.,  1996, Phys. Rev., D54, 2535

\bibitem[\protect\citeauthoryear{Martins \& Shellard}{Martins \&
  Shellard}{2002}]{martins2000}
Martins C. J. A.~P.,  Shellard E. P.~S.,  2002, Phys. Rev., D65, 043514

\bibitem[\protect\citeauthoryear{Mauskopf et~al.,}{Mauskopf
  et~al.}{2000}]{mauskopf2000}
Mauskopf P.~D.,  et~al., 2000, Astrophys. J., 536, L59

\bibitem[\protect\citeauthoryear{{Morales}}{{Morales}}{2006}]{morales2006}
{Morales} M.~F.,  2006, \apjl, 650, L21

\bibitem[\protect\citeauthoryear{{Morganti}, {Rottgering}, {Snellen}, {Miley},
  {Barthel}, {Best}, {Bruggen}, {Brunetti}, {Chyzy}, {Conway}, {Jarvis} \&
  {Lehnert}}{{Morganti} et~al.}{2010}]{morganti2010}
{Morganti} R.,  {Rottgering} H.,  {Snellen} I.,  {Miley} G.,  {Barthel} P.,
  {Best} P.,  {Bruggen} M.,  {Brunetti} G.,  {Chyzy} K.,  {Conway} J.,
  {Jarvis} M.,    {Lehnert} M.,  2010, ArXiv e-prints

\bibitem[\protect\citeauthoryear{Namikawa, Yamauchi \& Taruya}{Namikawa
  et~al.}{2012}]{namikawa2011}
Namikawa T.,  Yamauchi D.,    Taruya A.,  2012, JCAP, 1201, 007

\bibitem[\protect\citeauthoryear{Netterfield, Devlin, Jarosik, Page \&
  Wollack}{Netterfield et~al.}{1997}]{netterfield1996}
Netterfield C.~B.,  Devlin M.~J.,  Jarosik N.,  Page L.,    Wollack E.~J.,
  1997, Astrophys. J., 474, 47

\bibitem[\protect\citeauthoryear{{Patel}, {Bacon}, {Beswick}, {Muxlow} \&
  {Hoyle}}{{Patel} et~al.}{2010}]{patel2010}
{Patel} P.,  {Bacon} D.~J.,  {Beswick} R.~J.,  {Muxlow} T.~W.~B.,    {Hoyle}
  B.,  2010, \mnras, 401, 2572

\bibitem[\protect\citeauthoryear{{Pen} \& {Mao}}{{Pen} \&
  {Mao}}{2006}]{pen2006}
{Pen} U.-L.,  {Mao} S.,  2006, \mnras, 367, 1543

\bibitem[\protect\citeauthoryear{{Planck Collaboration}, {Ade}, {Aghanim},
  {Armitage-Caplan}, {Arnaud}, {Ashdown}, {Atrio-Barandela}, {Aumont},
  {Baccigalupi}, {Banday} \& et al.}{{Planck Collaboration}
  et~al.}{2014}]{planckstrings}
{Planck Collaboration} {Ade} P.~A.~R.,  {Aghanim} N.,  {Armitage-Caplan} C.,
  {Arnaud} M.,  {Ashdown} M.,  {Atrio-Barandela} F.,  {Aumont} J.,
  {Baccigalupi} C.,  {Banday} A.~J.,    et al. 2014, \aap, 571, A25

\bibitem[\protect\citeauthoryear{{Planck Collaboration}, {Ade}, {Aghanim},
  {Arnaud}, {Ashdown}, {Aumont}, {Baccigalupi}, {Banday}, {Barreiro},
  {Bartlett} \& et al.}{{Planck Collaboration} et~al.}{2015}]{planckparams2015}
{Planck Collaboration} {Ade} P.~A.~R.,  {Aghanim} N.,  {Arnaud} M.,  {Ashdown}
  M.,  {Aumont} J.,  {Baccigalupi} C.,  {Banday} A.~J.,  {Barreiro} R.~B.,
  {Bartlett} J.~G.,    et al. 2015, ArXiv e-prints

\bibitem[\protect\citeauthoryear{Prasanna \& Mohanty}{Prasanna \&
  Mohanty}{2002}]{prasanna2002}
Prasanna A.~R.,  Mohanty S.,  2002, EPL (Europhysics Letters), 60, 651

\bibitem[\protect\citeauthoryear{{Schmidt} \& {Jeong}}{{Schmidt} \&
  {Jeong}}{2012}]{schmidt2012}
{Schmidt} F.,  {Jeong} D.,  2012, \prd, 86, 083527

\bibitem[\protect\citeauthoryear{{Smail}, {Ellis} \& {Fitchett}}{{Smail}
  et~al.}{1994}]{smail1994}
{Smail} I.,  {Ellis} R.~S.,    {Fitchett} M.~J.,  1994, \mnras, 270, 245

\bibitem[\protect\citeauthoryear{{Smith}, {Peacock}, {Jenkins}, {White},
  {Frenk}, {Pearce}, {Thomas}, {Efstathiou} \& {Couchman}}{{Smith}
  et~al.}{2003}]{smith2003}
{Smith} R.~E.,  {Peacock} J.~A.,  {Jenkins} A.,  {White} S.~D.~M.,  {Frenk}
  C.~S.,  {Pearce} F.~R.,  {Thomas} P.~A.,  {Efstathiou} G.,    {Couchman}
  H.~M.~P.,  2003, \mnras, 341, 1311

\bibitem[\protect\citeauthoryear{{Stebbins}}{{Stebbins}}{1996}]{stebbins1996}
{Stebbins} A.,  1996, ArXiv Astrophysics e-prints

\bibitem[\protect\citeauthoryear{{Stil}, {Krause}, {Beck} \& {Taylor}}{{Stil}
  et~al.}{2009}]{stil2009}
{Stil} J.~M.,  {Krause} M.,  {Beck} R.,    {Taylor} A.~R.,  2009, \apj, 693,
  1392

\bibitem[\protect\citeauthoryear{{Surpi} \& {Harari}}{{Surpi} \&
  {Harari}}{1999}]{surpi1999}
{Surpi} G.~C.,  {Harari} D.~D.,  1999, \apj, 515, 455

\bibitem[\protect\citeauthoryear{{The Dark Energy Survey Collaboration}}{{The
  Dark Energy Survey Collaboration}}{2005}]{des}
{The Dark Energy Survey Collaboration} 2005, ArXiv Astrophysics e-prints

\bibitem[\protect\citeauthoryear{{Thomas}, {Bruni} \& {Wands}}{{Thomas}
  et~al.}{2015}]{thomas2015}
{Thomas} D.~B.,  {Bruni} M.,    {Wands} D.,  2015, \jcap, 9, 021

\bibitem[\protect\citeauthoryear{{Thomas}, {Contaldi} \& {Magueijo}}{{Thomas}
  et~al.}{2009}]{thomas2009}
{Thomas} D.~B.,  {Contaldi} C.~R.,    {Magueijo} J.,  2009, Physical Review
  Letters, 103, 181301

\bibitem[\protect\citeauthoryear{{Van Waerbeke}, {Mellier}, {Erben},
  {Cuillandre}, {Bernardeau}, {Maoli}, {Bertin}, {McCracken}, {Le F{\`e}vre},
  {Fort}, {Dantel-Fort}, {Jain} \& {Schneider}}{{Van Waerbeke}
  et~al.}{2000}]{vanwaerbeke2000}
{Van Waerbeke} L.,  {Mellier} Y.,  {Erben} T.,  {Cuillandre} J.~C.,
  {Bernardeau} F.,  {Maoli} R.,  {Bertin} E.,  {McCracken} H.~J.,  {Le
  F{\`e}vre} O.,  {Fort} B.,  {Dantel-Fort} M.,  {Jain} B.,    {Schneider} P.,
  2000, \aap, 358, 30

\bibitem[\protect\citeauthoryear{{Whittaker}, {Brown} \& {Battye}}{{Whittaker}
  et~al.}{2014}]{whittaker2014}
{Whittaker} L.,  {Brown} M.~L.,    {Battye} R.~A.,  2014, \mnras, 445, 1836

\bibitem[\protect\citeauthoryear{{Whittaker}, {Brown} \& {Battye}}{{Whittaker}
  et~al.}{2015}]{whittaker2015}
{Whittaker} L.,  {Brown} M.~L.,    {Battye} R.~A.,  2015, \mnras, 451, 383

\bibitem[\protect\citeauthoryear{{Wittman}, {Tyson}, {Kirkman}, {Dell'Antonio}
  \& {Bernstein}}{{Wittman} et~al.}{2000}]{wittman2000}
{Wittman} D.~M.,  {Tyson} J.~A.,  {Kirkman} D.,  {Dell'Antonio} I.,
  {Bernstein} G.,  2000, \nat, 405, 143

\bibitem[\protect\citeauthoryear{{Yamauchi}, {Namikawa} \& {Taruya}}{{Yamauchi}
  et~al.}{2012}]{yamauchi2012}
{Yamauchi} D.,  {Namikawa} T.,    {Taruya} A.,  2012, \jcap, 10, 030

\bibitem[\protect\citeauthoryear{Yamauchi, Namikawa \& Taruya}{Yamauchi
  et~al.}{2013}]{yamauchi2013}
Yamauchi D.,  Namikawa T.,    Taruya A.,  2013, JCAP, 1308, 051

\end{thebibliography}

\end{document}